\documentclass[prl,aps,onecolumn]{revtex4}
\usepackage{bm}
\usepackage{multirow}
\usepackage{ctable}
\usepackage[greek,dutch,english]{babel}
\usepackage{amsmath}                              
\usepackage{amssymb}                              
\usepackage[small,bf]{caption}                    
\usepackage{fancyhdr}                             
\usepackage{graphicx}                             
\usepackage{tabularx}                             
\usepackage{subfigure}                            
\usepackage{color}
\usepackage{natbib}
\usepackage{makeidx}

\usepackage{bbm}                                  
\usepackage{textcomp}                             

\newcommand{\ket}[1]{\ensuremath{|#1\rangle}}
\newcommand{\bra}[1]{\ensuremath{\langle#1|}}

\begin{document}

\title{Universal control and error correction in multi-qubit spin registers in diamond}

\author{T.~H.~Taminiau$^1$, J. Cramer$^1$, T. van der Sar$^1$, V. V. Dobrovitski$^2$, and R. Hanson$^1$}
\affiliation{$^1$Kavli Institute of Nanoscience, Delft University
of Technology, PO Box 5046, 2600 GA Delft, The Netherlands.\\
$^2$Ames Laboratory and Iowa State University, Ames, Iowa 50011,
USA.}

\begin{abstract}
Quantum registers of nuclear spins coupled to electron spins of
individual solid-state defects are a promising platform for
quantum information processing [1-13]. Pioneering experiments
selected defects with favourably located nuclear spins having
particularly strong hyperfine couplings [4-10]. For progress
towards large-scale applications, larger and deterministically
available nuclear registers are highly desirable. Here we realize
universal control over multi-qubit spin registers by harnessing
abundant weakly coupled nuclear spins. We use the electron spin of
a nitrogen-vacancy centre in diamond to selectively initialize,
control and read out carbon-13 spins in the surrounding spin bath
and construct high-fidelity single- and two-qubit gates. We
exploit these new capabilities to implement a three-qubit
quantum-error-correction protocol [14-17] and demonstrate the
robustness of the encoded state against applied errors. These
results transform weakly coupled nuclear spins from a source of
decoherence into a reliable resource, paving the way towards
extended quantum networks and surface-code quantum computing based
on multi-qubit nodes [11,18,19].
\end{abstract}

\maketitle

Electron and nuclear spins associated with defects in solids
provide natural hybrid quantum registers [3-11]. Fully-controlled
registers of multiple spins hold great promise as building blocks
for quantum networks [18] and fault-tolerant quantum computing
[19]. The defect electron spin enables initialization and readout
of the register and coupling to other (distant) electron spins
[11,18], whereas the nuclear spins provide well-isolated qubits
and memories with long coherence times [8,9,11]. Previous
experiments relied on selected defects having nuclear spins with
strong hyperfine couplings that exceed the inverse of the electron
spin dephasing time ($1/T_2^*$). With these strongly coupled
spins, single-shot readout [9,10,20-22] and entanglement [9,11]
were demonstrated. However, the number of strongly coupled spins
varies per defect and is intrinsically limited, so that universal
control has so far been restricted to two-qubit registers [4,7]
and the required control of multi-qubit registers has remained an
open challenge.

Here we overcome this challenge by demonstrating universal control
of weakly coupled nuclear spins (unresolved hyperfine coupling
$1/T_2^*$). We use the electron spin of single nitrogen-vacancy
(NV) centres in room-temperature diamond to selectively control
multiple carbon-13 ($^{13}$C) nuclear spins in the surrounding
spin bath (Fig. 1a). With this new level of control we realize
multi-qubit registers by constructing high-fidelity unconditional
and electron-controlled gates, implementing initialization and
readout, and creating nuclear-nuclear entangling gates through the
electron spin. Finally, we demonstrate the power of this approach
by implementing the first quantum-error-correction protocol with
individual solid-state spins.

\begin{figure}[t]
\begin{center}
\includegraphics[scale=0.70]{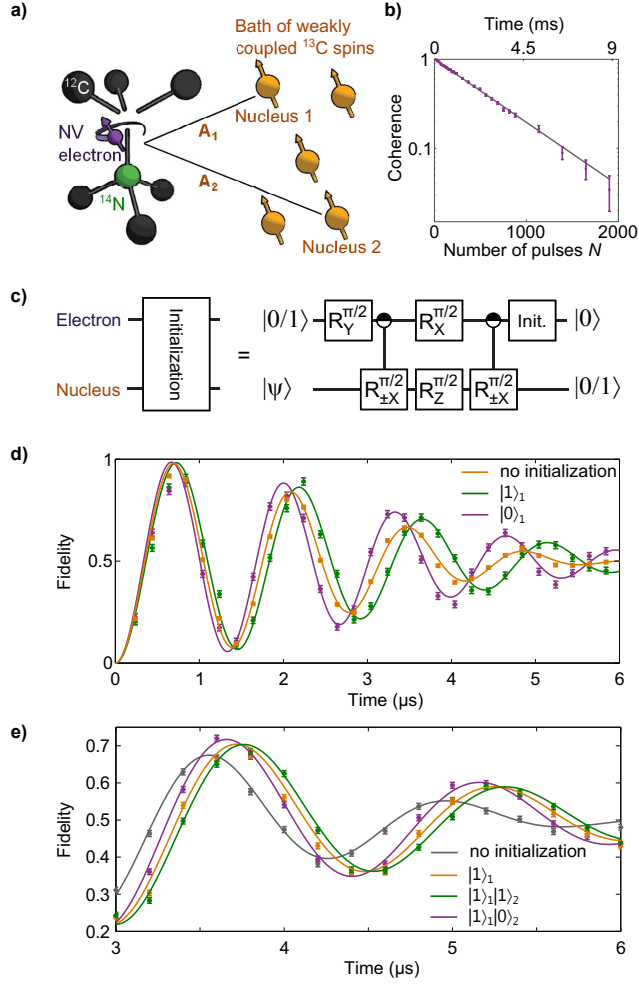}
\end{center}
\caption{\textbf{Definition and initialization of the quantum
registers.} (a) Quantum register formed by the nitrogen-vacancy
(NV) electron spin ($S=1$; $\ket{0} = \ket{m_s=0}$,$\ket{1} =
\ket{m_s=-1}$) and weakly coupled $^{13}$C nuclear spins ($I
=1/2$; state $\ket{\psi}_i$ and hyperfine interaction $A_i$ for
nuclear spin $i$, see Methods for values). All gates on nuclear
spins are implemented by sequences of $N$ pi-pulses on the
electron spin spaced by a time $2\tau$ (Methods). (b) The
electronic coherence as a function of the total sequence length.
The number of pi-pulses $N$ is increased for fixed
$\tau=2\pi/\omega_L$, which is representative for our gates.
$\omega_L = 2\pi \cdot 431$ kHz is the $^{13}$C Larmor frequency.
The 1/e time is $T_{coh}=2.86(4)$ ms. (c) Nuclear spin
initialization by swapping the electron state, $\ket{0/1} =
\ket{0}$ or $\ket{1}$, onto the nuclear spin. The controlled gates
($R_{\pm X}^{\pi/2}$) are $X$-rotations by $\pi/2$ with a
direction conditional on the electron spin state (Methods). The
final electron spin re-initialization by a $2\ \mu$s laser pulse
(labelled ``Init.'') preserves the nuclear spin polarization
($T_1$ values under illumination: $2.5(3)$ ms for nuclear spin 1
and $1.2(2)$ ms for nuclear spin 2, Supplementary Information).
(d) Electron Ramsey measurements without nuclear spin
initialization and with nuclear spin 1 initialized in $\ket{0}_1$
or $\ket{1}_1$, and (e) with nuclear spin 1 initialized in
$\ket{1}_1$  and nuclear spin 2 in $\ket{0}_2$ or $\ket{1}_2$. All
error bars and uncertainties in this work are $1\sigma$.}
\label{Figure1} \vspace*{-0.2cm}
\end{figure}

We have used dynamical decoupling spectroscopy [23-25] to
characterize the nuclear spin environment of a total of three NV
centres, including one with an additional strongly coupled
$^{13}$C spin (Supplementary Information). To demonstrate the
universality of our approach to create multi-qubit registers, we
have realized initialization, control and readout of three weakly
coupled $^{13}$C spins for each NV centre studied (Supplementary
Information). Below we consider one of these NV centres in detail
and use two of its weakly coupled $^{13}$C spins to form a
three-qubit register for quantum error correction (Fig. 1a).

Our control method exploits the dependence of the nuclear spin
quantization axis on the electron spin state due to the
anisotropic hyperfine interaction (see Methods for hyperfine
parameters), so that no radio-frequency driving of the nuclear
spins is required [23-27]. All nuclear gates are implemented by
pulse sequences of the form ($\tau - \pi - 2\tau - \pi -
\tau)^{N/2}$ where $\pi$ is a microwave pi-pulse on the electron
spin, $2\tau$ is the inter-pulse delay and $N$ is the total number
of pulses in the sequence. Each nuclear spin is controlled by
precisely choosing $\tau$ in resonance with that spin's particular
hyperfine interaction. The target spin, the type of gate
(conditional or unconditional) and the rotation axis ($X$- or
$Z$-rotation) are determined by the value of $\tau$; the total
rotation angle is determined by $N$ (Methods). Crucially, these
sequences at the same time decouple the electron from the other
nuclear qubits and the environment [7]; these
decoherence-protected gates are selective and allow the full
electron coherence time $T_{coh}$ to be exploited
($T_{coh}=2.86(4)$ ms, Fig. 1b). The gates are thus not limited by
the electron spin dephasing time $T_2^*=3.3(1)\ \mu$s or Hahn echo
time $T_2$ and do not require strong coupling.

To initialize the nuclear spins we first prepare the electron spin
in $m_s=0$ by optical pumping (Supplementary Information), then
swap the electron state onto the nuclear spin, and finally
re-initialize the electron spin (Fig. 1c).  We characterize the
nuclear initialization by preparing the electron spin in a
superposition state and letting it evolve in a Ramsey-type
experiment. Without initialization a single-frequency oscillation
with a Gaussian decaying envelope is observed, confirming that the
NV centre feels a decohering bath of weakly coupled spins (Fig.
1d). Initializing one of the nuclear spins in the $\ket{0}$
($\ket{1}$) state (Fig. 1d and 1e), we increase (decrease) the
oscillation frequency because the magnetic field at the electron
is enhanced (reduced) due to the hyperfine interaction. The
oscillations also persist longer as quasistatic fluctuations of
the two nuclear spins are suppressed [28], increasing the
electronic dephasing time to $T_2^*=4.0(2)\ \mu$s. From this data,
we obtain state initialization fidelities of $F_1  = 0.91(2)$ and
$F_2 = 0.88(5)$ for nuclear spin 1 and 2 respectively (see
Methods).

\begin{figure}[t]
\begin{center}
\includegraphics[scale=0.70]{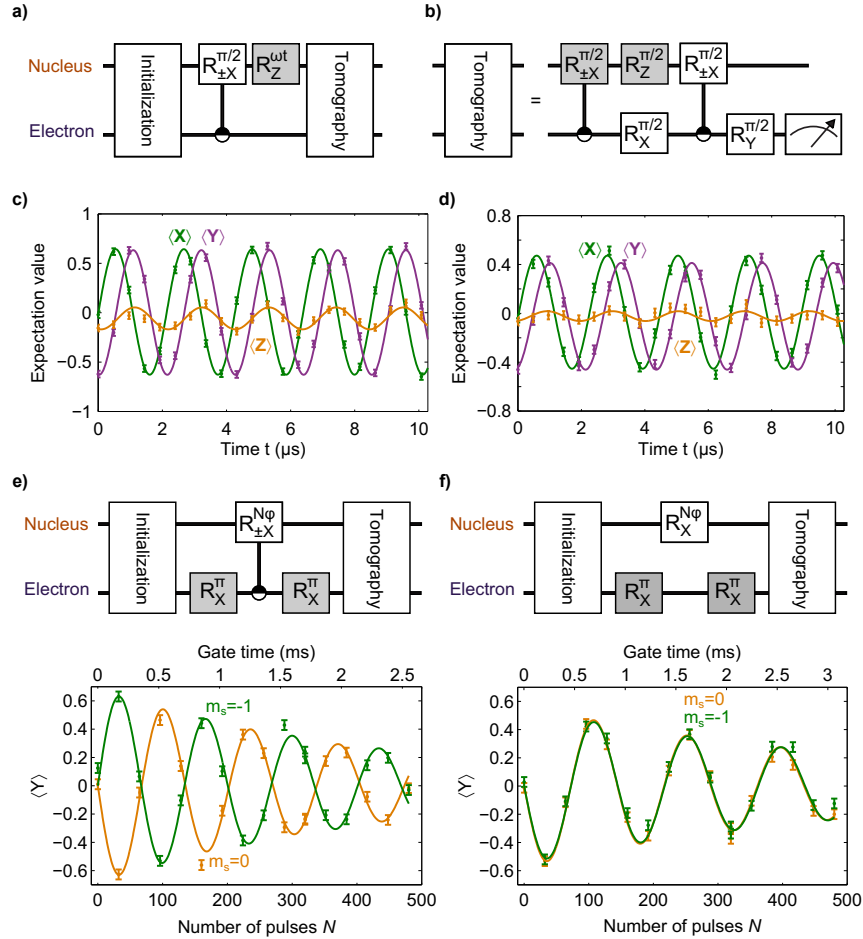}
\end{center}
\caption{\textbf{Individual nuclear spin control and readout.} (a)
Sequence for nuclear-spin free-precession experiments. The
$Z$-rotation is implemented by an off-resonant sequence ($\tau =
2\pi/\omega_L$) with a variable number of pulses $N$. (b) Nuclear
spin state tomography is performed by mapping the
$\langle{X}\rangle$, $\langle{Y}\rangle$ and $\langle{Z}\rangle$
expectation values onto the electron spin and reading out the
electron (shaded gates are optional basis rotations). (c-d)
Measurement of $\langle{X}\rangle$, $\langle{Y}\rangle$ and
$\langle{Z}\rangle$ as function of the free-evolution time. The
oscillations in $\langle{X}\rangle$ and $\langle{Y}\rangle$
confirm the selective control and readout of the targeted nuclear
spins. The amplitude yields a combined readout and initialization
fidelity of $0.82(1)$ for spin 1 in (c) and $0.72(1)$ for spin 2
in (d). Curves are sinusoidal fits. See Supplementary Information
for a complete data set with three nuclear spins for each of the
three NV centres studied, demonstrating the universality of the
control method. (e) Characterization of the conditional gate for
nuclear spin 1. The nuclear spin rotates about $X$ with opposite
directions for $m_s=0$ (without shaded gates) and $m_s=-1$ (with
shaded gates). Time for a $\pm \pi/2$-rotation: $170\ \mu$s. (f)
Unconditional gate for nucleus 1; the rotation is independent of
the electron state. Time for a $\pi/2$-rotation: $254\ \mu$s. See
Supplementary Information for gates on nuclear spin 2. Results are
not corrected for initialization or readout fidelities.}
\label{Figure2} \vspace*{-0.2cm}
\end{figure}

Next we demonstrate the measurement of the individual nuclear spin
states and verify that we observe two distinct $^{13}$C spins by
performing nuclear free-evolution experiments (Fig. 2a-d). The
oscillations in the expectation values $\langle{X}\rangle$ and
$\langle{Y}\rangle$ show that the nuclear spins states are
successfully read out. The precession frequencies, $\omega=2\pi
 \cdot 470(1)$ kHz for nuclear spin 1 (Fig. 2c) and $\omega=2\pi
 \cdot 449(2)$ kHz for nuclear spin 2 (Fig. 2d), are different and
agree with the average of $\omega = \omega_L$ (for $m_s=0$) and
$\omega \approx \omega_L+ A_\parallel$ (for $m_s=-1$), as expected
because the electron spin is continuously flipped. $A_\parallel$
is the parallel component of the hyperfine interaction (Methods)
and $\omega_L = 2\pi \cdot 431$ kHz is the bare nuclear Larmor
frequency. These results confirm that we selectively address the
two targeted $^{13}$C spins.

Universal control requires both conditional and unconditional
gates, while maintaining a high degree of coherence for all qubits
in the register. To characterize our gates, we initialize the
nuclear spins, prepare the electron spin either in $m_s=0$ or in
$m_s=-1$ and apply a gate with a variable number of pulses. For
the conditional gate, $\langle{Y}\rangle$ oscillates in anti-phase
for the two electron states: the nuclear spin rotates around $X$
in a direction that depends on the initial electron state (Fig.
2e). In contrast, for the unconditional gate the rotation
direction is independent of the electron state (Fig. 2f). The slow
decay of the oscillations indicates that high gate fidelities are
possible ($F \sim 0.96$ for a nuclear $\pi$/2-rotation), enabling
us to explore multi-gate sequences that implement nuclear-nuclear
gates and quantum error correction.

\begin{figure}[t]
\begin{center}
\includegraphics[scale=0.70]{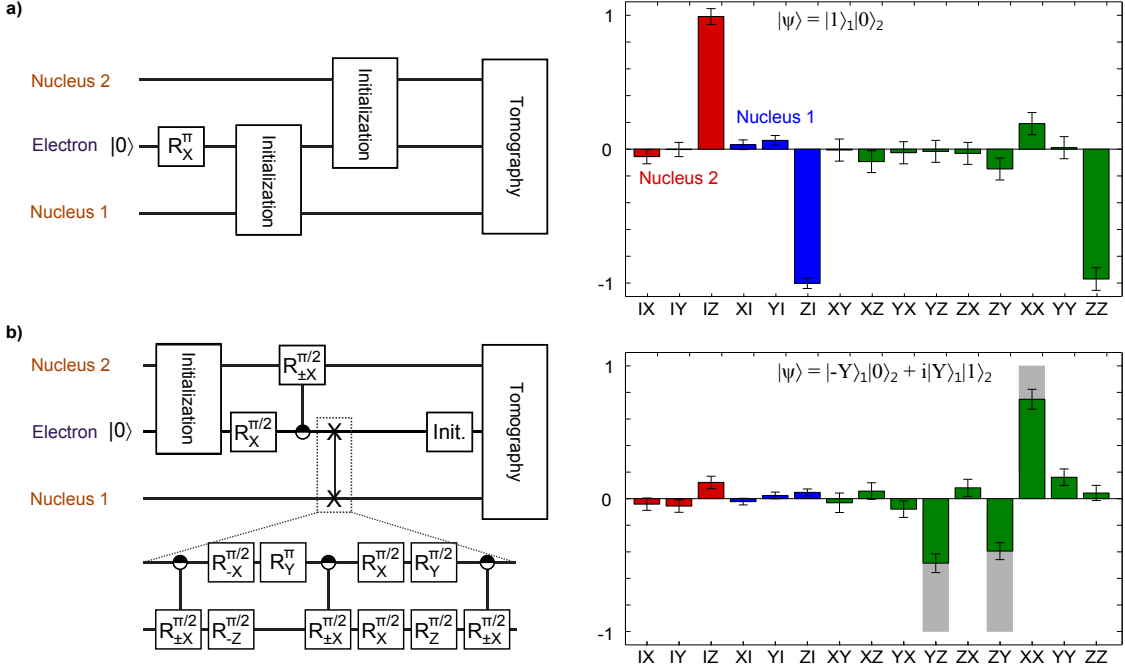}
\end{center}
\caption{\textbf{Two-qubit control and nuclear-nuclear entangling
gate.} (a) The nuclear spins are prepared in $\ket{1}_1\ket{0}_2$
and two-qubit tomography is performed by mapping the $15$
combinations of Pauli operators $\langle{\sigma_i\sigma_j}\rangle$
onto the electron spin (Supplementary Information). After
correction for single-qubit initialization and readout fidelities,
the state fidelity with the target state is $F=0.99(3)$,
indicating that the sequential initialization and two-qubit
readout are accurate. (b) Entangling gate between nuclear spins by
coherently swapping the state of the electron onto nuclear spin 1.
The gate consists of $167$ electron operations over $986\ \mu$s
(excluding initialization and tomography). The nuclear spin
coherence is preserved during electron spin re-initialization (a
$2\ \mu$s laser pulse); $T_2^*$ values under illumination are
$51(7)\ \mu$s and $0.35(9)$ ms for nuclear spin 1 and 2
respectively (Supplementary Information). The grey bars depict the
target state.} \label{Figure3} \vspace*{-0.2cm}
\end{figure}

To realize quantum gates between the nuclear spins [27,29], whose
mutual interaction is negligible, we use the electron spin as a
quantum bus. We first verify that both nuclear spins can be
prepared and read out in the same experiment by initializing the
spins in an eigenstate and performing state tomography by mapping
the two-qubit correlations onto the electron spin (Fig. 3a). We
then implement entangling gates through an electron controlled
gate on nuclear spin 2 and a subsequent coherent SWAP gate between
the electron and nuclear spin 1 (Fig. 3b). The tomography reveals
strong correlations between the nuclear spins with near-zero
single-qubit expectation values, a clear signature of an
entangling gate. The fidelity with the target state is $0.66(3)$
(initialization and readout corrected), demonstrating that the
gate can take a pure input state into an entangled state of
nuclear spins.

Finally, we implement a quantum-error-correction protocol that
protects a quantum state from bit-flip errors by encoding it in a
$3$-qubit state and correcting errors through majority voting
(Fig. 4a). Such protocols have been realized with nuclear magnetic
resonance [14,15], trapped ions [16] and superconducting qubits
[17], but have so far been out of reach for individual solid-state
spins due to a lack of multi-qubit control. We compose this
protocol from one- and two-qubit gates (Fig. 4b) and separately
confirm that the constructed doubly-controlled gate flips the
state around X only if the control qubits (nuclear spins) are in
$\ket{1}_1\ket{1}_2$ (Fig. 4c).

\begin{figure}[t]
\begin{center}
\includegraphics[scale=0.70]{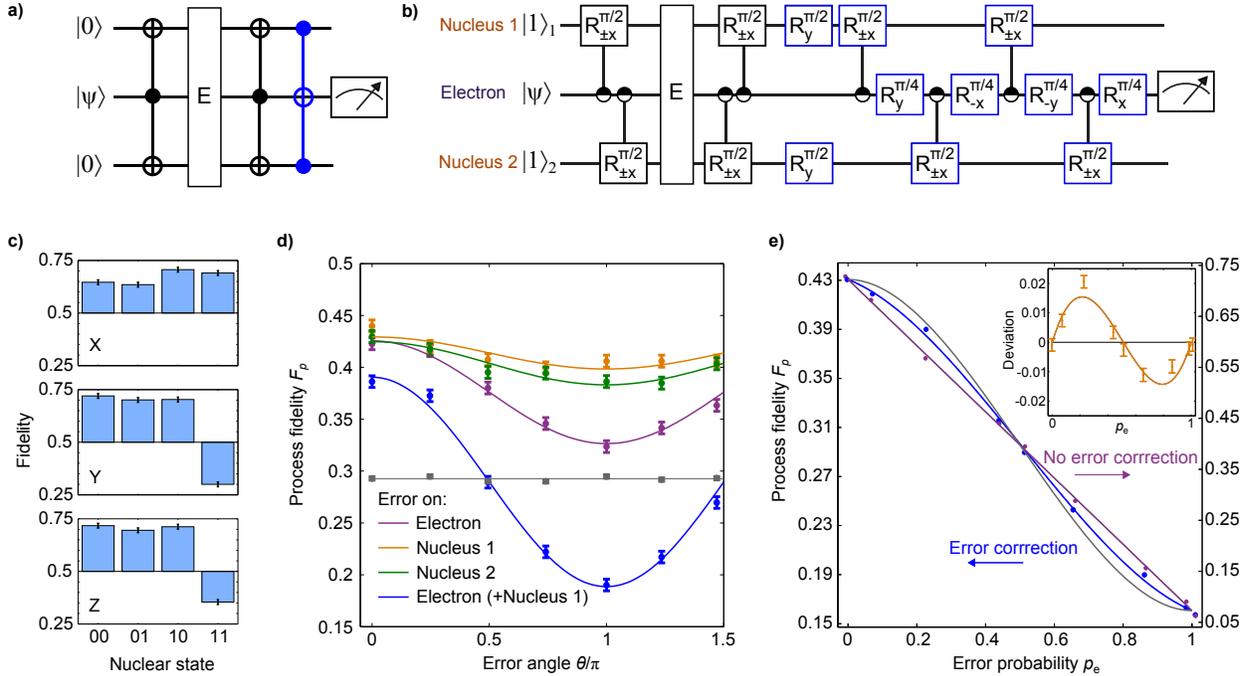}
\end{center}
\caption{\textbf{Implementation of 3-qubit quantum error
correction.} (a) Bit-flip quantum-error-correction protocol. The
state $\ket{\psi}$ is encoded in an entangled state using two
ancilla qubits. Potential errors $E$ are detected by decoding and
are corrected based on a doubly-controlled NOT gate. (b) Our
implementation of the quantum-error-correction protocol in (a).
The doubly-controlled gate (blue) is constructed using 4
controlled gates as the final ancilla states are irrelevant. The
experiment consists of $308$ electron operations in $1.8$ ms
(excluding initialization). (c) Characterization of the
doubly-controlled gate (blue gates in (b) only). The average
output fidelities for $\ket{\pm X}$, for $\ket{\pm Y}$ and for
$\ket{\pm Z}$ are shown for the four ancilla basis states. The
average process fidelity with the targeted action is $F_p  =
0.534(5)$. (d) Process fidelity for errors applied to nucleus 1,
to nucleus 2, or to the electron spin (with and without additional
flip of nuclear spin 1).  Grey data and fit are $F_{p0}=(F_x+
F_{-x})/4$ which sets the average value for the expected
oscillations if no errors are corrected. (e) Process fidelity for
errors simultaneously applied to all three qubits with error
probability $p_e$. Purple: without error correction. Blue: with
error correction. Grey: for ideal robustness against errors. Error
bars are given by the symbol size (typical standard deviation
$0.002$). Inset: deviation of the error correction data from a
linear curve. All curves in (d) and (e) are fits to the model in
the Methods section.} \label{Figure4} \vspace*{-0.2cm}
\end{figure}

We first characterize the effect of errors on each individual
qubit. The applied errors are rotations around the X-axis by an
angle $\theta$ with a random sign ($50\%$ clockwise, $50\%$
anticlockwise) and therefore represent a decoherence-type process
with a strength determined by $\theta$. We prepare $6$ input
states $\ket{\psi} = \ket{\pm X}, \ket{\pm Y}$ and $\ket{\pm Z}$,
measure the corresponding fidelities $F$ of the output states and
calculate the process fidelity $F_p$  with the identity process:

\begin{equation}\nonumber
F_p = \frac{F_x+F_{-x}+F_y+F_{-y}+F_z+F_{-z}}{4}-1/2
\end{equation}

Without error correction, errors on the data qubit (electron spin)
are expected to result in an oscillation about $F_{p0} = (F_x+
F_{-x})/4$ because only the $\ket{\pm X}$ states are unaffected by
the applied errors. Instead, the experimental process fidelity
with error correction always remains above $F_{p0}$, even for a
completely randomizing error ($\theta = \pi/2$), indicating that
the state is partly recovered (Fig. 4d). If one of the ancilla
qubits (nuclear spins) is also flipped, an oscillation about
$F_{p0}$ is observed; the error correction is effectively turned
off because the protocol cannot correct two-qubit errors.

To quantitatively determine the effectiveness of the error
correction we analyze it in terms of the three probabilities $p_n$
that an applied error on qubit $n$ is successfully corrected and a
decoherence/depolarization process during the error-correction
protocol itself (Methods). The model accurately fits the data and
gives $p_1  = 0.63(1)$, $p_2  = 0.89(2)$ and $p_3  = 0.84(2)$ for
errors on the electron, nucleus 1 and nucleus 2 respectively.
Crucially, the average probability $\langle{p_n}\rangle =
(p_1+p_2+p_3)/3 = 0.786(9)$ is well above $2/3$, demonstrating
that the process is robust against applied single-qubit errors and
that the entropy associated with the errors is successfully
shuttled to the ancilla qubits.

We further demonstrate the robustness by applying errors
simultaneously on all three qubits (Fig. 4e). Without error
correction, i.e. without doubly-controlled gate, a linear
dependence is observed and a fit to the expected form gives
$\langle{p_n}\rangle=0.67(3)$ in excellent agreement with
$\langle{p_n}\rangle=2/3$ expected for no robustness to errors.
With error correction a markedly slower initial decay and a
non-linear behaviour with $\langle{p_n}\rangle=0.84(3)$ is
obtained. This suppression of the linear dependence is a key
characteristic of quantum error correction.

The deviation from $\langle{p_n}\rangle=1$ is mainly due to
imperfect nuclear initialization, which might be improved by
repeated initialization steps (Supplementary Information) or
projective measurements [9,21]. We calculate
$\langle{p_n}\rangle=0.94(2)$ for ideal initialization fidelity
(Supplementary Information). Without applied errors, decoherence
and depolarization during the protocol itself (more than $300$
electron operations over $1.8$ ms) result in a process fidelity of
$F_p=0.431(2)$, corresponding to an average gate fidelity of
$0.93$ for the $10$ nuclear spin gates. The main source of
infidelity is electron decoherence ($T_{coh}  = 2.86(4)$ ms, Fig.
1b), which is likely phonon-induced [30] and limits the average
gate fidelity to $0.97$. Nuclear spin dephasing further reduces
the fidelity to $0.94$, close to the observed value (Supplementary
Information). The electronic coherence time is greatly increased
at cryogenic temperatures, at which $T_{coh}= 14$ ms (single NV)
[18] and $T_{coh}  = 0.6$ s (ensembles) [30] have already been
reported. Nuclear spin dephasing can be mitigated by decoupling
nuclear-nuclear interactions ($T_2$ measurements in Supplementary
Information). With such future improvements, our results can be
used to protect entangled states of solid-state spins.

In conclusion, we have established universal control over weakly
coupled nuclear spins that were previously regarded as a source of
decoherence. These results provide multiple qubits per defect with
high certainty and are compatible with control of the intrinsic
nitrogen spin and potential strongly coupled $^{13}$C spins. Our
techniques can be applied to a wide variety of other
electron-nuclear spin systems [2,3,10,13]. The resulting reliable
multi-qubit registers can be combined with recently demonstrated
coherent coupling between (distant) electron spins [11,18] to
realize novel surface-code quantum-computation architectures that
use four qubits per defect node [19] and extended quantum networks
for long-distance quantum communication.

\section{\textbf{Methods}}

\subsection{Diamond sample and hyperfine interactions}

We use a room-temperature type IIa diamond with $1.1\%$ of
$^{13}$C grown by chemical vapor deposition (Element 6). We apply
a magnetic field of $B_z \approx 403$ G along the NV symmetry axis
($Z$-axis), yielding a $^{13}$C Larmor frequency $\omega_L  = 2\pi
\cdot 431$ kHz. The electronic dephasing time $T_2^*$ is $3.3(1)\
\mu$s. The hyperfine interaction for nuclear spin $i$ is given by
$A_i= A_\parallel^i\hat{z}+ A_\perp^i\hat{x}$ (Fig. 1a), with
$A_\parallel$ the component parallel to the magnetic field and
$A_\perp$ the perpendicular component. $A_\parallel^1 = 2\pi \cdot
78.2(8)$ kHz and $A_\perp^1 = 2\pi \cdot 30(1)$ kHz for nuclear
spin 1, and $A_\parallel^2 = 2\pi \cdot 32(3)$ kHz and $A_\perp^2
= 2\pi \cdot 44(2)$ kHz for nuclear spin 2. Because
$A_\parallel,A_\perp < (2\sqrt{2})/T_2^* =2\pi \cdot 136(1)$ kHz
the nuclear spins are weakly coupled to the electron spin and the
hyperfine splittings are unresolved.

\subsection{Nuclear gate design}

In a suitable rotating frame, the Hamiltonian with a single
nuclear spin can be written:

\begin{equation}\nonumber
\hat{H} =  \ket{0}\bra{0}\hat{H}_0 + \ket{1}\bra{1}\hat{H}_1
\end{equation}

with $\hat{H}_0 = \omega_L \hat{I}_Z$  and $\hat{H}_1 = (\omega_L
+ A_\parallel)\hat{I}_Z + A_\perp\hat{I}_X$ and with $\ket{0}$ and
$\ket{1}$ the $m_s=0$ and $m_s=-1$ electron states, respectively.
Nuclear spin gates are performed by applying sequences of the type
$(\tau-\pi-2\tau-\pi-\tau)^{(N/2)}$ on the electron spin (Rabi
frequency $31.25$ MHz). Because we set $\omega_L \gg A_\perp$,
sharp resonances occur at $\tau \approx \frac{k\pi}{2\omega_L+
A_\parallel}$, with integer $k$. At these values a nuclear
$X$-rotation is performed (assuming $A_\perp \neq 0$). For odd $k$
the direction of the rotation is conditional on the electron spin
(e.g. the $R_{\pm X}^{\pi/2}$ gates), for even $k$ it is
unconditional ($R_{X}^{\pi/2}$ gates). For the conditional gates
we use $\tau=2.656\ \mu$s, $N=32$ for spin 1 and $\tau = 3.900\
\mu$s, $N=18$ for spin 2. For the unconditional gates we use
$\tau=3.186\ \mu$s, $N=40$ for spin 1 and $\tau = 2.228\ \mu$s,
$N=64$ for spin 2. Z-rotations are implemented by choosing $\tau$
off-resonant. Detailed simulations of the nuclear spin dynamics
are available in the Supplementary Information.

\subsection{Nuclear spin initialization fidelity}

The electron Ramsey measurements in Fig. 1d and 1e are analyzed in
two ways: (1) The measurements are separately fit to $F=1/2-1/2
e^{-(t/T_2^*)^2}\cos(\omega t)$, in which $T_2^*$ is a measure for
the dephasing time set by the entire spin bath. The external
magnetic field stability of better than $2$ mG over the total
integration time ($\sim2$ hours), required in these experiments,
was achieved by post selecting from a larger measurement set. (2)
We determine the nuclear spin initialization fidelities $F_1$ and
$F_2$ by averaging over multiple measurement runs (Supplementary
Information) and using the hyperfine components $A_\parallel^1$
and $A_\parallel^2$ together with:

\begin{equation}\nonumber
F = 1/2 - 1/2 e^{-(t/T_2^{**})^2}\Bigg(F_1 F_2
\cos\Bigg[\left(\omega+\frac{A_\parallel^1+ A_\parallel^2}{2}\right)t\Bigg]
+ F_1(1-F_2)
\cos\Bigg[\left(\omega+\frac{A_\parallel^1-A_\parallel^2}{2}\right)t\Bigg]
\end{equation}

\begin{equation}\nonumber
+ (1-F_1)F_2\cos\Bigg[\left(\omega+\frac{-A_\parallel^1+
A_\parallel^2}{2}\right)t\Bigg]+(1-F_1)(1-F_2)\cos\Bigg[\left(\omega-\frac{A_\parallel^1+
A_\parallel^2}{2}\right)t\Bigg]\Bigg).
\end{equation}
Here $T_2^{**}=4.5(3)\ \mu$s is the electronic dephasing due to
the rest of the spin bath, i.e. not including the two spins under
study.

\subsection{Quantum error correction analysis}

The applied errors realize the quantum map:

\begin{equation}\nonumber
E(\rho,\theta) = \cos^2(\theta/2) I\rho{I} + \sin^2(\theta/2)
X\rho{X},
\end{equation}
in which $\rho$ is the initial density matrix (error
characterization in Supplementary Information). We analyze the
error-correction protocol by separating depolarization during the
encoding, decoding and error-correction steps from the robustness
of the encoded state to applied errors, which is characterized by
the three probabilities $p_n$ that an error applied on qubit $n$
is successfully corrected (derivation in Supplementary
Information). The process fidelity for a single-qubit error (Fig.
4d) is then given by:

\begin{equation}\nonumber
F_p(\theta)= F_{p0} + A_{YZ}[p_n+(1-p_n)\cos\theta],
\end{equation}
where $F_{p0} = (F_x+ F_{-x})/4$ and $A_{YZ}=(F_y + F_{-y} + F_z +
F_{-z}-2)/4$ characterize the additional depolarization and are
given by the average fidelities without applied errors. The
equation contains a constant due to the $\ket{\pm X}$ states,
which are unaffected by the applied error, and a sum of successful
($p_n=1$) and unsuccessful ($p_n=0$) error correction for the
$\ket{\pm Y}$ and $\ket{\pm Z}$ states. For errors simultaneous on
all three qubits (Fig. 4e), the process fidelity becomes:

\begin{equation}\nonumber
F_p(p_e) = F_{p0} + A_{YZ} [1 - 3p_e + 3p_e^2 - 2p_e^3 + 3(2
\langle{p_n}\rangle-1)(p_e-3p_e^2+2p_e^3)],
\end{equation}
with $p_e=\sin^2(\theta/2)$ the error probability. In general this
equation describes a third order polynomial. For ideal error
correction ($\langle{p_n}\rangle=1$) the linear term vanishes,
whereas without robustness to errors ($\langle{p_n}\rangle=2/3$),
the result is strictly linear. The inversion symmetry about
$p_e=0.5$ observed both theoretically and experimentally ensures
that the nonlinear behavior is not due to spurious coherent
rotations.

\section{Note}

After submission of this manuscript we became aware of related work by Waldherr et al., arXiv:1309.6424 in which 3-qubit quantum error correction is implemented using strongly coupled nuclear spins. 

\section{Acknowledgements}

We thank L. Childress, J. J. L. Morton, O. Moussa and L.M.K.
Vandersypen for helpful discussions and comments. T.H.T.
acknowledges support by a Marie Curie Intra European Fellowship
within the 7th European Community Framework Programme. Work at the
Ames Laboratory was supported by the U.S. Department of Energy
Basic Energy Sciences under contract no. DE-AC02-07CH11358. We
acknowledge support from the Dutch Organization for Fundamental
Research on Matter (FOM), the Netherlands Organization for
Scientific Research (NWO), the DARPA QuASAR programme, the EU
SOLID, and DIAMANT programmes and the European Research Council
through a Starting Grant.

\section{References}

\noindent [1] Awschalom, D.D., Bassett, L.C., Dzurak, A.S., Hu,
E.L., \& Petta, J.R., Quantum Spintronics: Engineering and
Manipulating Atom-Like Spins in Semiconductors. \textit{Science}
\textbf{339}, 1174-1179 (2013).

\noindent [2] Koehl, W.F., Buckley, B.B., Heremans, F.J.,
Calusine, G., \& Awschalom, D.D., Room temperature coherent
control of defect spin qubits in silicon carbide. \textit{Nature}
\textbf{479}, 84-87 (2011).

\noindent [3] Yin, C. et al., Optical addressing of an individual
erbium ion in silicon. \textit{Nature} \textbf{497}, 91-94 (2013).

\noindent [4] Gurudev Dutt, M.V. et al., Quantum Register Based on
Individual Electronic and Nuclear Spin Qubits in Diamond.
\textit{Science} \textbf{316}, 1312-1316 (2007).

\noindent [5] Neumann, P. et al., Multipartite Entanglement Among
Single Spins in Diamond. \textit{Science} \textbf{320}, 1326-1329
(2008).

\noindent [6] Fuchs, G.D., Burkard, G., Klimov, P.V., \&
Awschalom, D.D., A quantum memory intrinsic to single
nitrogen-vacancy centres in diamond. \textit{Nature Phys.}
\textbf{7}, 789-793 (2011).

\noindent [7] van der Sar, T. et al., Decoherence-protected
quantum gates for a hybrid solid-state spin register.
\textit{Nature} \textbf{484}, 82-86 (2012).

\noindent [8] Maurer, P.C. et al., Room-Temperature Quantum Bit
Memory Exceeding One Second. \textit{Science} \textbf{336},
1283-1286 (2012).

\noindent [9] Pfaff, W. et al., Demonstration of
entanglement-by-measurement of solid-state qubits. \textit{Nature
Phys.} \textbf{9}, 29-33 (2013).

\noindent [10] Pla, J.J. et al., High-fidelity readout and control
of a nuclear spin qubit in silicon. \textit{Nature} \textbf{496},
334-338 (2013).

\noindent [11] Dolde, F. et al., Room-temperature entanglement
between single defect spins in diamond. \textit{Nature Phys.}
\textbf{9}, 139-143 (2013).

\noindent [12] Jiang, L. et al., Repetitive Readout of a Single
Electronic Spin via Quantum Logic with Nuclear Spin Ancillae.
\textit{Science} \textbf{326}, 267-272 (2009).

\noindent [13] Lee, S.-Y. et al., Readout and control of a single
nuclear spin with a metastable electron spin ancilla.
\textit{Nature Nanotech.} \textbf{8}, 487-492 (2013).

\noindent [14] Cory, D.G. et al., Experimental Quantum Error
Correction. \textit{Phys. Rev. Lett.} \textbf{81}, 2152-2155
(1998).

\noindent [15] Moussa, O., Baugh, J., Ryan, C.A., \& Laflamme, R.,
Demonstration of Sufficient Control for Two Rounds of Quantum
Error Correction in a Solid State Ensemble Quantum Information
Processor. \textit{Phys. Rev. Lett.} \textbf{107}, 160501 (2011).

\noindent [16] Schindler, P. et al., Experimental Repetitive
Quantum Error Correction. \textit{Science} \textbf{332}, 1059-1061
(2011).

\noindent [17] Reed, M.D. et al., Realization of three-qubit
quantum error correction with superconducting circuits.
\textit{Nature} \textbf{482}, 382-385 (2012).

\noindent [18] Bernien, H. et al., Heralded entanglement between
solid-state qubits separated by three metres. \textit{Nature}
\textbf{497}, 86-90 (2013).

\noindent [19] Nickerson, N.H., Li, Y., \& Benjamin, S.C.,
Topological quantum computing with a very noisy network and local
error rates approaching one percent. \textit{Nat. Commun.}
\textbf{4}, 1756 (2013).

\noindent [20] Neumann, P. et al., Single-Shot Readout of a Single
Nuclear Spin. \textit{Science} \textbf{329}, 542-544 (2010).

\noindent [21] Robledo, L. et al., High-fidelity projective
read-out of a solid-state spin quantum register. \textit{Nature}
\textbf{477}, 574-578 (2011).

\noindent [22] Dreau, A., Spinicelli, P., Maze, J.R., Roch, J.F.,
\& Jacques, V., Single-Shot Readout of Multiple Nuclear Spin
Qubits in Diamond under Ambient Conditions. \textit{Phys. Rev.
Lett.} \textbf{110}, 060502 (2013).

\noindent [23] Taminiau, T.H. et al., Detection and Control of
Individual Nuclear Spins Using a Weakly Coupled Electron Spin.
\textit{Phys. Rev. Lett.} \textbf{109}, 137602 (2012).

\noindent [24] Kolkowitz, S., Unterreithmeier, Q.P., Bennett,
S.D., \& Lukin, M.D., Sensing Distant Nuclear Spins with a Single
Electron Spin. \textit{Phys. Rev. Lett.} \textbf{109}, 137601
(2012).

\noindent [25] Zhao, N. et al., Sensing single remote nuclear
spins. \textit{Nature Nanotech.} \textbf{7}, 657-662 (2012).

\noindent [26] Hodges, J.S., Yang, J.C., Ramanathan, C., \& Cory,
D.G., Universal control of nuclear spins via anisotropic hyperfine
interactions. \textit{Phys. Rev. A} \textbf{78}, 010303 (2008).

\noindent [27] Zhang, Y., Ryan, C.A., Laflamme, R., \& Baugh, J.,
Coherent Control of Two Nuclear Spins Using the Anisotropic
Hyperfine Interaction. \textit{Phys. Rev. Lett.} \textbf{107},
170503 (2011).

\noindent [28] London, P. et al., Detecting and polarizing nuclear
spins with double resonance on a single electron spin.
\textit{Phys. Rev. Lett.} \textbf{111}, 067601 (2013).

\noindent [29] Filidou, V. et al., Ultrafast entangling gates
between nuclear spins using photoexcited triplet states.
\textit{Nature Phys.} \textbf{8}, 596-600 (2012).

\noindent [30] Bar-Gill, N., Pham, L.M., Jarmola, A., Budker, D.,
\& Walsworth, R.L., Solid-state electronic spin coherence time
approaching one second. \textit{Nat. Commun.} \textbf{4}, 1743
(2013).

\newpage

\section{\textbf{Supplementary Material}}


\section{Setup and sample}

The experimental setup and sample are described in detail in the
supplementary information of Van der Sar et al. \cite{Sar2012} We
used a type-IIa chemical vapour deposition grown diamond with a
$1.1\%$ natural abundance of carbon-13 (Element 6). Solid
immersion lenses were fabricated on top of the nitrogen vacancy
(NV) centres to enhance the collection efficiency
\cite{Bernien2012}. The electron spin is controlled by microwaves
through an on-chip stripline (Rabi frequency of 31.25 MHz). A
magnetic field of $B_z \approx 403$ G was applied along the NV
symmetry axis using three electromagnets. At this magnetic field
the intrinsic NV nitrogen-14 spin is polarized due to an
excited-state anti-crossing \cite{Smeltzer2009,Jacques2009}.

\section{Electron spin initialization and readout}

This section discusses the electron spin initialization,
re-initialization and readout. In particular it analyzes how the
imperfect spin and charge state initialization affect the outcomes
of the different type of experiments performed.
%

\subsection{Experimental}

The electron spin is initialized in the $m_s=0$ state by a $532$
nm ($\sim 150 \mu$W) laser pulse (typically 4$\mu$s) and read out
through its spin-dependent time-resolved fluorescence. In all
experiments we measure the difference signal $\Delta_f = S_f -
\tilde{S_f}$ between the fluorescence signal $S_f$ for the final
state and the fluorescence signal $\tilde{S}_f$ for the final
state with a pi-pulse applied just before readout ($m_s=0$ to
$m_s=-1$ transition). The obtained value is then normalized by
dividing it by the same difference signal right after
initialization: $\Delta_i = S_i - \tilde{S_i}$, where $S_i$ is
without pi-pulse and $\tilde{S}_i$ with pi-pulse. The final
normalized contrast $C$ is:

\begin{equation}
C = \frac{S_f - \tilde{S}_f}{S_i - \tilde{S}_i} =
\frac{\Delta_f}{\Delta_i}.
\end{equation}
This method directly measures the contrast between $m_s=0$ and
$m_s=-1$ states. Note that $-1\leq C\leq1$ and that the result is
independent of the population in other states, such as $m_s=+1$,
that are not affected by the microwave pi-pulse. The reported
expectation values directly correspond to $C$, the measured
fidelities are obtained from $F = C/2+1/2$.

\subsection{Initial electron state}

The electronic initialization involves both spin states
($m_s=-1,0,+1$) and charge states ($NV^-$ and $NV^0$). The initial
state $\rho_{i}$ is:

\begin{equation}\label{Eq: electron init}
\rho_{i} = p_{1}\rho_0 + p_2\rho_{m} + p_3\rho_{s} + p_4\rho_{c},
\end{equation}
with $p_1+p_2+p_3+p_4 = 1$, and in which $\rho_0$ is the desired
$m_s=0$ state, $\rho_{m}$ is the completely mixed state of $m_s=0$
and $m_s=-1$, $\rho_{s}$ represents the other spins states (here
$m_s=+1$) and $\rho_{c}$ other charge states (here $NV^0$).

The precise values for $p_1, p_2, p_3$ and $p_4$ are unknown. For
this NV centre the spin-state initialization fidelity was
previously reported to be $F_s = \frac{p_{1}+p_2/2}{p_1+p_2+p_3} >
0.95$ under similar conditions \cite{Sar2012}. The NV$^-$
charge-state initialization fidelity $F_c = p_1+p_2+p_3$ is
unknown here, but values of $\sim 0.7$ have been reported for
other NV centres \cite{Aslam2013}.

The available initial population is given by $p_1$. Ideally,
measurements of $C = \frac{\Delta_f}{\Delta_i}$ directly reflect
the actual polarization so that $-p_1 \leq C \leq p_1$. Because
only the $\rho_0$ term in equation \ref{Eq: electron init} is
affected by microwave pulses only this term yields signal
(non-zero $\Delta$), so that the normalization signal always is
$\Delta_i = D_0p_{1}$, with $D_0$ an unknown proportionality
constant. Next we determine the obtainable final signal $\Delta_f$
for two types of experiments: experiments that do not
re-initialize the electron to create additional polarization in
the nuclear spin register and those that do.

\subsection{Experiments without nuclear spin initialization}

First consider experiments where only a single electron
initialization step is used, i.e. experiments that do not transfer
polarization to the nuclear spins before resetting the electron
spin. In this case the maximum value of $\Delta_f$ simply is
$D_0p_{1}$ and the maximum contrast is $C_{max} =
\frac{D_0p_{1}}{D_0p_{1}} = 1$. Due to the calibration, the final
measured contrast is independent of $p_1$ and therefore does not
take into account the charge and spin initialization fidelities.

\subsection{Experiments with nuclear spin initialization}

Now consider experiments in which the electron spin polarization
is transferred to a nuclear spin, the electron is re-initialized,
and finally the electron is used to measure the nuclear spin
state. The result $\Delta_f$ depends on the correlations of the
spin (charge) state after the re-initialization step with the spin
(charge) state before it. We assume the spin states before and
after re-initialization are uncorrelated, and derive the result
$\Delta_f$ for both uncorrelated (no memory) and maximally
positive correlations (ideal memory) for the charge state.

The state of the initialized electron and a single nuclear spin in
a completely mixed state is:
\begin{equation}\label{Eq: electron init_plusC13}
\rho = \rho_{electron} \otimes \rho_{nucleus} = p_{1}(\rho_0
\otimes \rho_m) + p_2(\rho_{m} \otimes \rho_{m}) + p_3(\rho_{s}
\otimes \rho_{m}) + p_4(\rho_{c} \otimes \rho_{m}),
\end{equation}
swapping the electron and nuclear spin states gives:
\begin{equation}\label{Eq: after_SWAP}
\rho = p_{1}(\rho_m \otimes \rho_0) + p_2(\rho_{m} \otimes
\rho_{m}) + p_3(\rho_{s} \otimes \rho_{m}) + p_4(\rho_{c} \otimes
\rho_{m}),
\end{equation}
as the SWAP gate has no effect on the erroneous electron spin
($\rho_s$) and charge ($\rho_c$) states. The electron spin
initialization $p_1$ is thus directly transferred to the nuclear
spin.

We re-initialize the electron spin and assume that electron spin
initialization is independent of the nuclear spin state. First
consider the case of no correlations (no memory) for the charge
state, so that the electron is completely re-initialized. The
state in equation \ref{Eq: after_SWAP} becomes:
\begin{equation}\label{Eq: electron init_plusC13}
\rho = (p_{1}\rho_0 + p_2\rho_{m} + p_3\rho_{s} + p_4\rho_{c})
\otimes (p_{1}\rho_0 + (1-p_1)\rho_m).
\end{equation}
Reading out the nuclear spin with the electron spin only yields
non-zero signal for both the electron and nuclear spins in the
pure state $\rho_0$, so that:

\begin{equation}\label{Eq: electron init_plusC13}
\Delta_f = D_0p_1^2,\ \ \  C_{max} = p_1,
\end{equation}
which shows that the maximum contrast $C_{max}$ is reduced by a
factor $p_1$ and thus that the experiment faithfully reflects the
actual nuclear spin state, including a reduced fidelity due to the
imperfect electron spin and charge initialization.

If the electron re-initialization does not change the charge
state, equation \ref{Eq: after_SWAP} after electron
re-initialization becomes:
\begin{equation}\label{Eq: electron init_plusC13}
\rho = p_1 \left(\frac{p_1\rho_0 + p_2\rho_m + p_3\rho_{s}}{p_1 +
p_2 + p_3}\right) \otimes \rho_0 + (p_2+p_3)\left(\frac{p_1\rho_0
+ p_2\rho_m + p_3\rho_{s}}{p_1 + p_2 + p_3}\right) \otimes \rho_m
+ p_4\rho_c \otimes \rho_m
\end{equation}
Again taking into account that no difference signal is obtained if
either the electron or the nuclear spin is not in state $\rho_0$:
\begin{equation}\label{Eq: electron init_plusC13}
\Delta_f = \frac{p_1^2}{p_1+p_2+p_3},\ \ \ C_{max} =
\frac{p_1}{p_1+p_2+p_3}
\end{equation}
The result now accurately reflects the spin state initialization,
but is independent of the charge state initialization.

The high nuclear initialization fidelity obtained here ($F \approx
0.9$, figure 1 of the main text), indicates that the charge state
initialization fidelity is high ($>0.90$) or that the measurements
are not sensitive to it (i.e. the re-initialization laser pulse
has low probability to change the charge state). The same value
gives a lower limit of the electron spin initialization $F = p_1/2
+1/2 \geq 0.90$, as the swap gate for initialization and the
nuclear spin readout have limited fidelities as well.

\subsection{Conclusion}

As in previous room temperature experiments, the charge state is
thus not rigourously initialized nor proven to be fully reflected
in the measurement outcomes. Therefore the measured state
fidelities do not give the actual purity of the states and no
entanglement can be proven to be present. Nevertheless the
(entangling) gates and protocols developed and studied in this
work can be accurately investigated through their action on the
prepared states. Note that methods to initialize the charge state
have been developed at room temperature \cite{Aslam2013} and that
pure entangled states have been reported at cryogenic temperatures
using simultaneous spin and charge initialization
\cite{Pfaff2013}.

\section{Characterization and control for three NV centres}

To demonstrate that harnessing weakly coupled spins makes multiple
qubits available for each defect with high certainty, we have
controlled three weakly coupled nuclear spins for each of the
three NV centres studied. This section contains the
characterization of the NV centres and the nuclear-spin
free-evolution experiments that demonstrate the initialization,
control and readout of the nuclear spins.

\subsection{Characterization of the nuclear spin environment}

\begin{figure}[t]
\begin{center}
\includegraphics[scale=0.70]{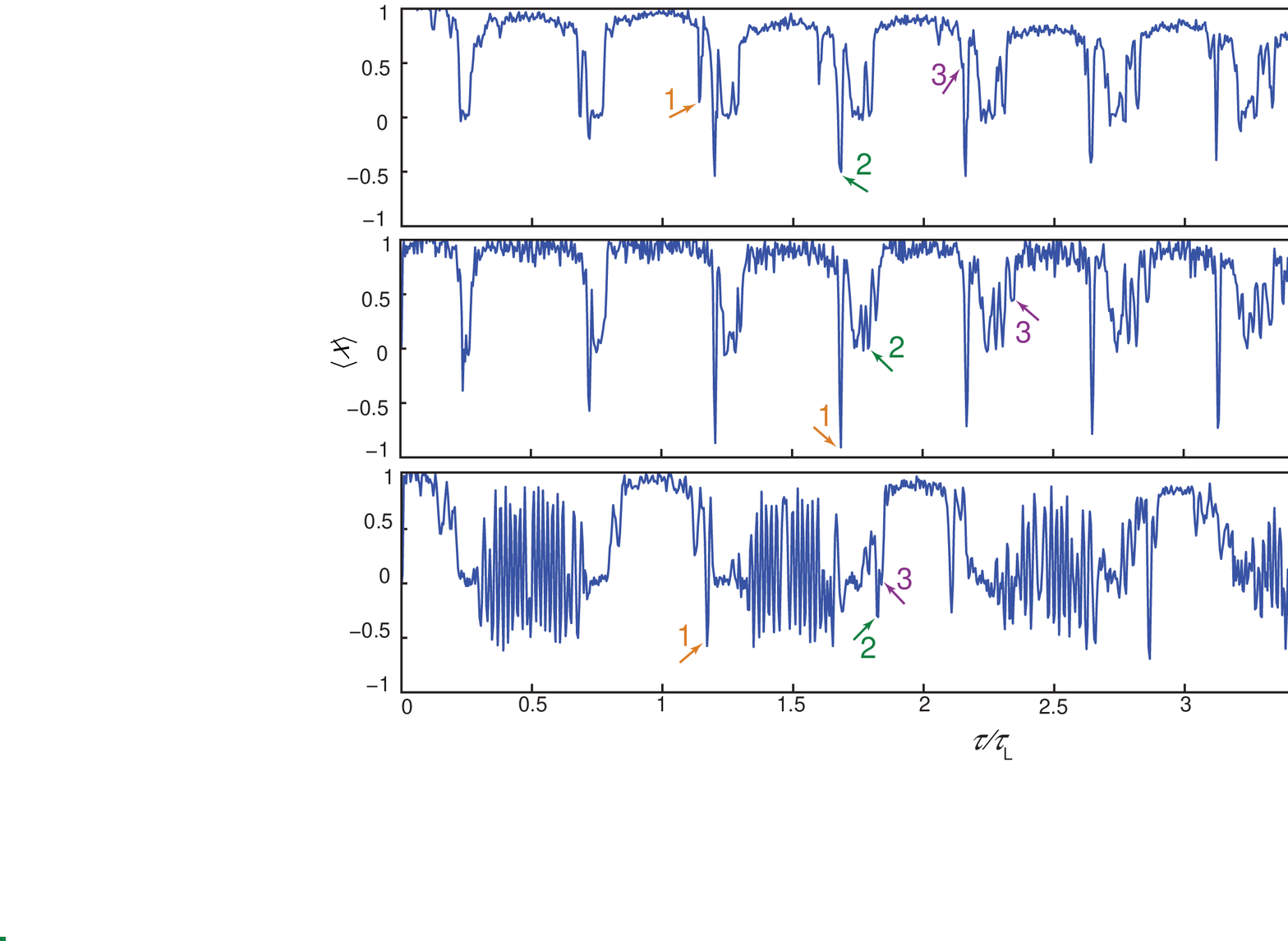}
\end{center}
\caption{\textbf{Characterization of the nuclear spin environment
for the three NV centres studied.} Dynamical decoupling
spectroscopy \cite{Taminiau2012} for NV$_A$, NV$_B$ and NV$_C$.
The electron spin is prepared in a coherent superposition state
($\ket{X} = \ket{0}+\ket{1}$) and a dynamical decoupling sequence
with 32 pi-pulses of the form $(\tau-\pi-2\tau-\pi-\tau)^{16}$ is
applied with variable interpulse delay $2\tau$ before measuring
$\langle X \rangle$. Sharp dips in the signal indicate an
entangling operation of the electron spin with individual $^{13}$C
spins in the spin bath \cite{Taminiau2012}. The arrows indicate
the 9 different $^{13}$C nuclear spins, and the values of $\tau$
used, for which we implemented initialization, control and readout
(see figures \ref{Fig:Ramseys_NV_A} for NV$_A$,
\ref{Fig:Ramseys_NV_B} for NV$_B$ and \ref{Fig:Ramseys_NV_C} for
NV$_C$). The experiments in the main text use nuclear spin 1 and 2
of NV$_A$. $\tau_L$ is the bare Larmor period.}
\label{Fig:Compare_NV} \vspace*{-0.2cm}
\end{figure}

We use dynamical decoupling spectroscopy \cite{Taminiau2012} to
characterize the nuclear spin environment of a total of three NV
centres: NV$_A$, which is studied in the main text, and the two
additional centres NV$_B$ and NV$_C$ (Fig. \ref{Fig:Compare_NV}).
The resulting curves provide characteristic fingerprints of the
nuclear spin environments of the NV centre.

NV$_A$ and NV$_B$ show qualitatively similar behavior (Fig.
\ref{Fig:Compare_NV}); both curves display broad echo collapses
due to the spin bath at $\tau/(4\tau_L) = m$ with odd $m$ and show
distinct sharp dips due to individual $^{13}$C nuclear spins that
become visible at larger $\tau$ \cite{Taminiau2012}. However, the
positions and depths of the different dips differ strongly due to
the characteristic distribution of nuclear spins near each NV
centre. In addition to a bath of weakly-coupled $^{13}$C spins
NV$_C$ shows a rapidly oscillating component in the signal due to
the presence of a strongly coupled nuclear spin (hyperfine
interaction of $2\pi\cdot 453$ kHz).

\subsection{Control of 3 weakly coupled nuclear spins per NV centre}

\begin{figure}[t]
\begin{center}
\includegraphics[scale=0.85]{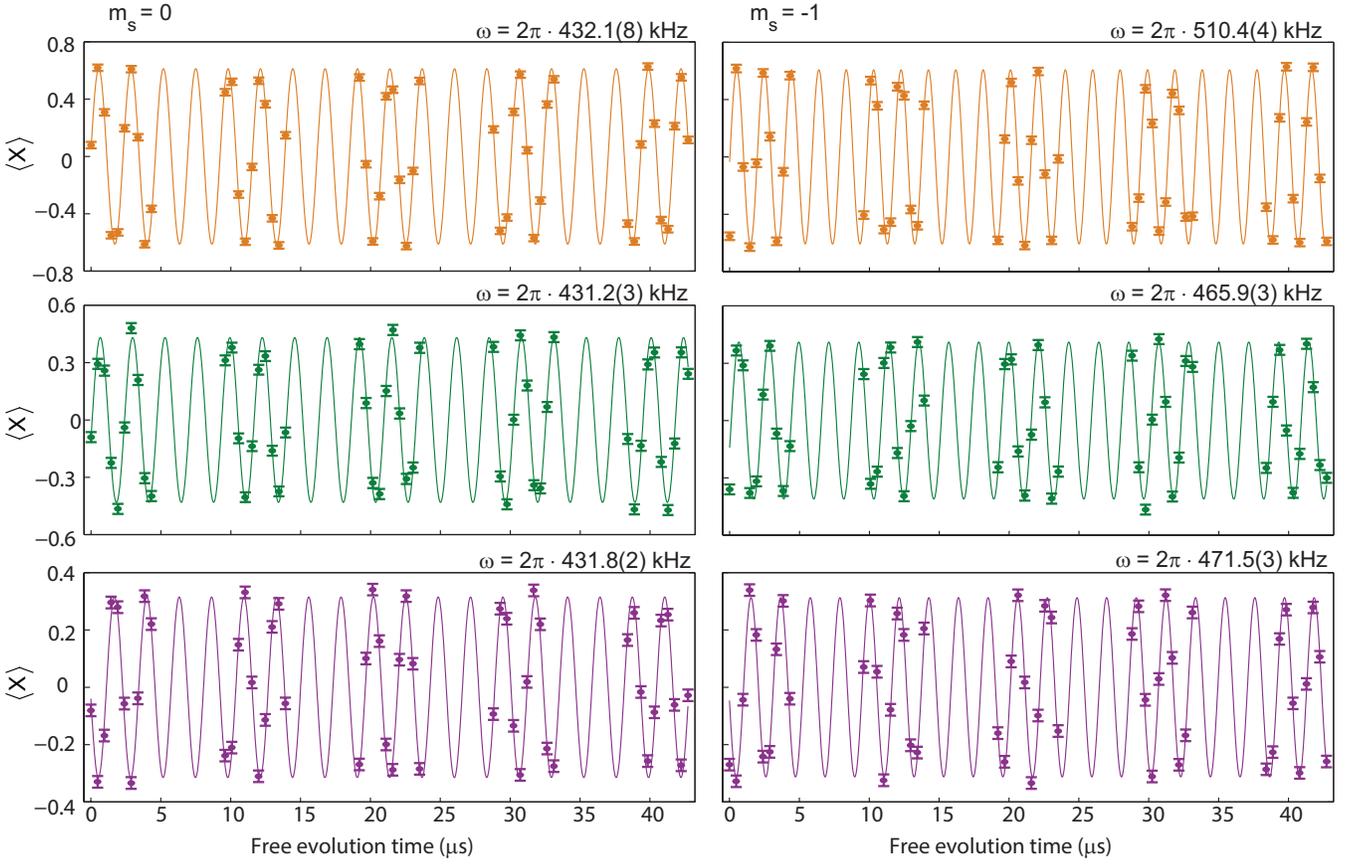}
\end{center}
\caption{\textbf{Control of three weakly coupled nuclear spins for
$\bf{NV_A}$}. Experiments as in figure 2 of the main text, but
with the electron spin in $m_s=0$ (left) or $m_s=-1$ (right). The
values for $\tau$ used for spin 1 (orange), spin 2 (green) and
spin 3 (purple) are marked in Fig. \ref{Fig:Compare_NV}. Note that
nuclear spin 1 and nuclear spin 2 are the two spins studied in
detail and used for implementing the quantum-error-correction
protocol.} \label{Fig:Ramseys_NV_A} \vspace*{-0.2cm}
\end{figure}

\begin{figure}[t]
\begin{center}
\includegraphics[scale=0.85]{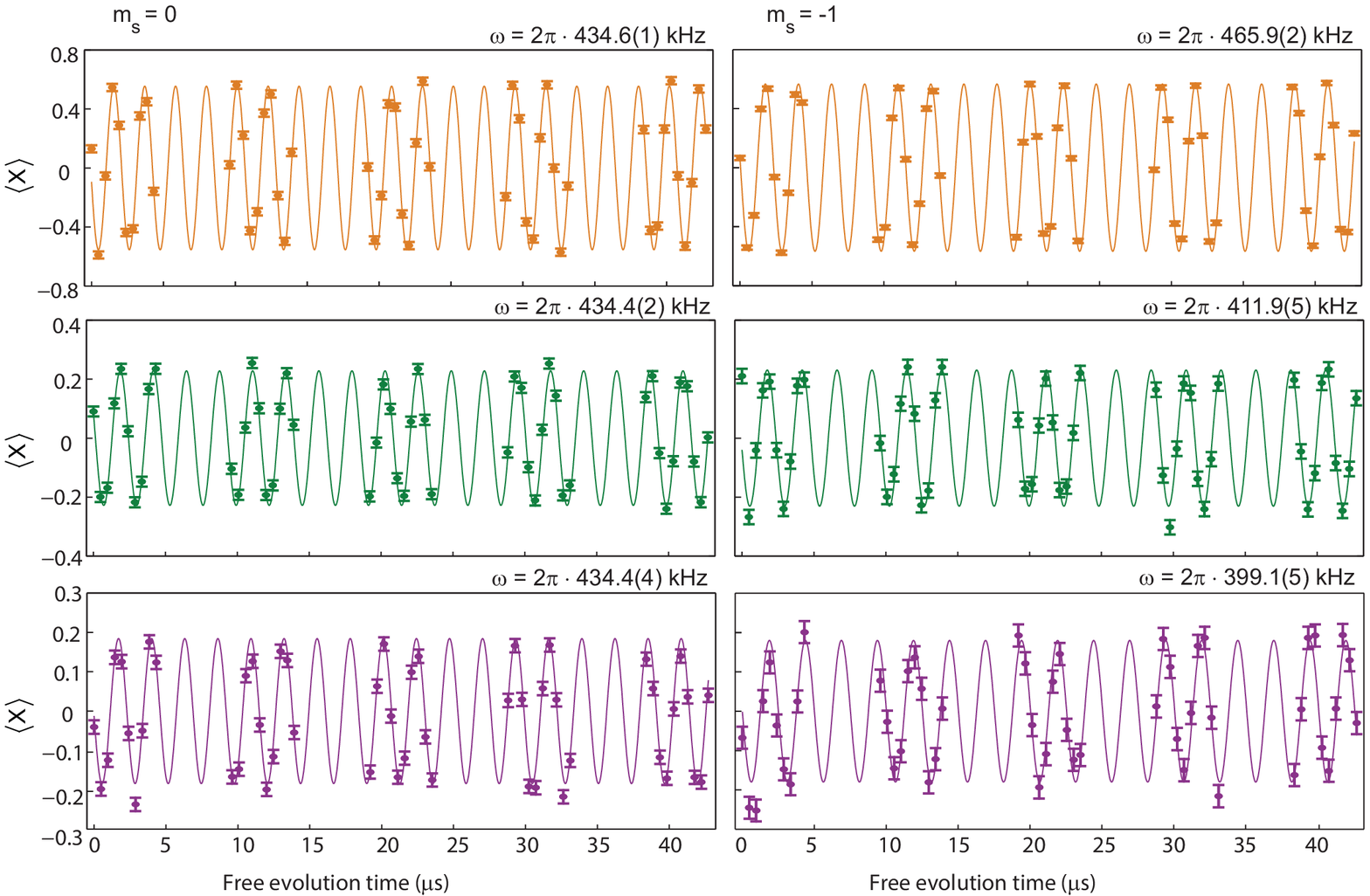}
\end{center}
\caption{\textbf{Control of three weakly coupled nuclear spins for
$\bf{NV_B}$}. Experiments as in figure 2 of the main text, but
with the electron spin in $m_s=0$ (left) or $m_s=-1$ (right). The
values for $\tau$ used for spin 1 (orange), spin 2 (green) and
spin 3 (purple) are marked in Fig. \ref{Fig:Compare_NV}.}
\label{Fig:Ramseys_NV_B} \vspace*{-0.2cm}
\end{figure}

\begin{figure}[t]
\begin{center}
\includegraphics[scale=0.85]{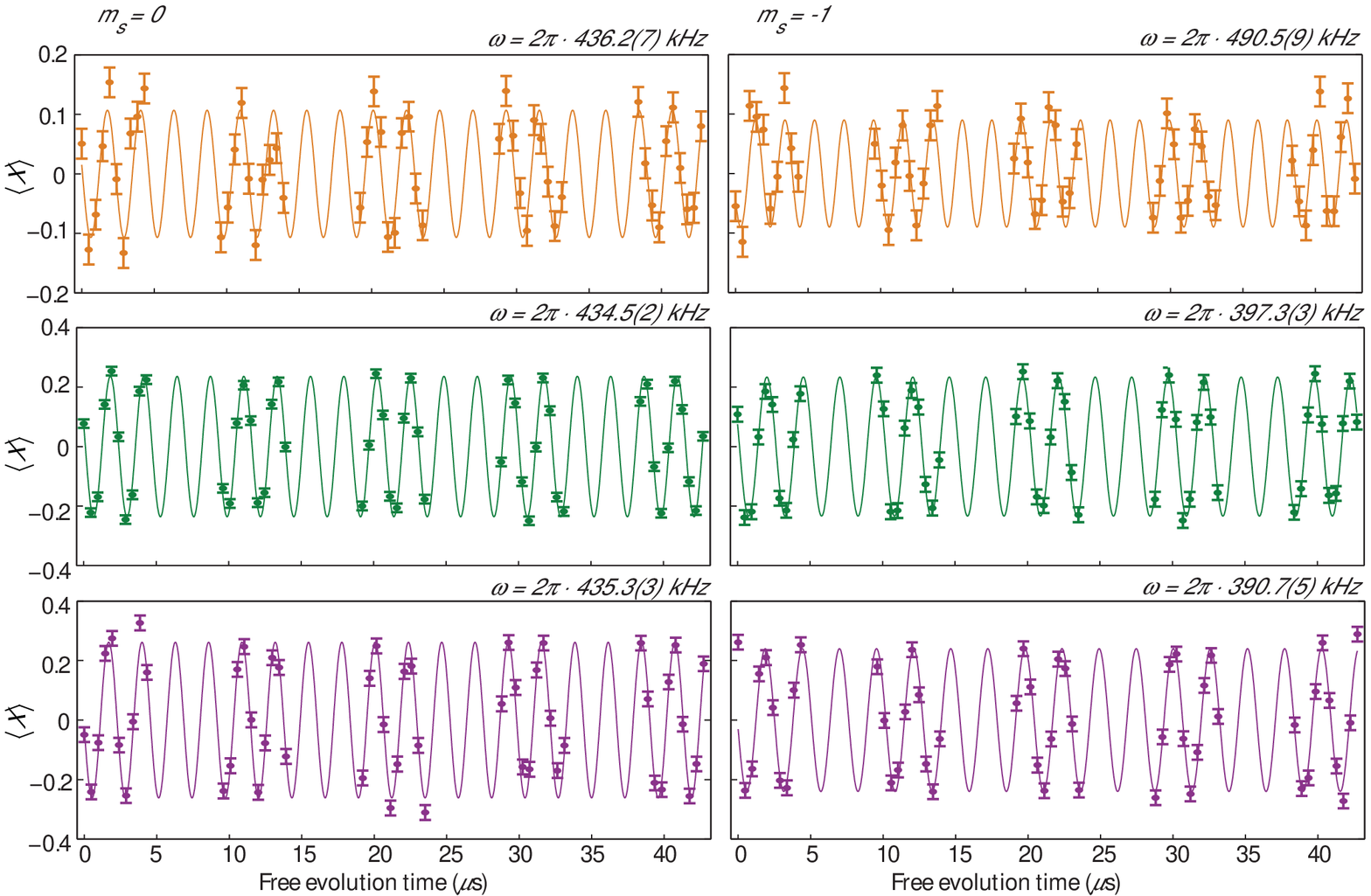}
\end{center}
\caption{\textbf{Control of three weakly coupled nuclear spins for
$\bf{NV_C}$}. Experiments as in figure 2 of the main text, but
with the electron spin in $m_s=0$ (left) or $m_s=-1$ (right). The
values for $\tau$ used for spin 1 (orange), spin 2 (green) and
spin 3 (purple) are marked in Fig. \ref{Fig:Compare_NV}.}
\label{Fig:Ramseys_NV_C} \vspace*{-0.2cm}
\end{figure}

For each of the three NV centres in figure \ref{Fig:Compare_NV},
we select three nuclear spins (marked in the figure) and
demonstrate initialization, control and direct readout by
performing nuclear free precession experiments (see figures 2a-d
of the main text). We prepare the electron spin in $m_s=0$ or
$m_s=-1$. The unique precession frequencies for $m_s=-1$ confirm
that in each case three different $^{13}$C spins are controlled
(Fig. \ref{Fig:Ramseys_NV_A}, \ref{Fig:Ramseys_NV_B} and
\ref{Fig:Ramseys_NV_C}).

These results demonstrate the control of three weakly-coupled
nuclear spins for each NV centre studied. Our
decoherence-protected gates therefore make several nuclear spins
available per defect centre with a high certainty, in stark
contrast to the highly probabilistic nature of the presence of
strongly coupled $^{13}$C spins. The fact that the gates can be
applied in the presence of strongly coupled nuclear spins,
including the intrinsic Nitrogen and nearby $^{13}$C nuclear
spins, indicates that quantum registers with over $5$ nuclear
spins are now well within reach (see e.g. NV$_C$ in figure
\ref{Fig:Compare_NV}).

\section{Nuclear spin dynamics and gates}

The hyperfine parameters for the three nuclear spins for NV$_A$,
the NV centre used in the main text, are given in Table
\ref{Table:Hyperfine_Parameters}. The two nuclear spin qubits in
the main text are spin 1 and 2.

\begin{table}[h]
  \centering
   \begin{tabular}{c|c|c}
      Nuclear spin  & Parallel component $A_\parallel$ (kHz)  & Perpendicular component $A_\perp$ (kHz) \\[1mm]
      \hline
      1 & $\ \ $78.2(8)$\ \ $ & $\ \ $30(1)$\ \ $ \\
      2 & $\ \ $32(3)$\ \ $ & $\ \   $44(2)$\ \ $ \\
      3 & $\ \ $41.2(4)$\ \ $ & $\ \ $19.2(7)$\ \ $ \\
   \end{tabular}
  \caption{\textbf{Hyperfine parameters for 3 relevant $^{13}$C spins.} $A_\parallel$ is the component parallel to the applied magnetic field (along the NV symmetry axis).
  $A_\perp$ is the perpendicular component. This NV centre was studied previously \cite{Taminiau2012}.}\label{Table:Hyperfine_Parameters}
\end{table}

\subsection{Nuclear gate design}

With an appropriate rotation of the coordinate axes, the
Hamiltonian of the NV electron spin and a single $^{13}$C spin is:
\begin{equation}
\hat{H}= A_\parallel {\hat S_z}  \hat{I_z} + A_\perp \hat{S_z}
\hat{I_x} + \omega_L \hat{I_z} = |0\rangle\langle 0| \hat{H_0} +
|1\rangle\langle 1| \hat{H_1},
\end{equation}
where $\hat{S_i}$ ($\hat{I_i}$) are the electron (nuclear) spin
operators, $\omega_L = 2\pi\cdot 431$ kHz is the nuclear Larmor
frequency (applied magnetic field $B_z \approx 403$G). The nuclear
spin evolution thus depends on the electron spin state:
$\hat{H_0}$ if the electron is in $m_s=0$ (state $|0\rangle$), and
$\hat{H_1}$ if the electron is in $m_s=-1$ (state $|1\rangle$),
with
\begin{equation}
\hat{H_0}=\omega_L \hat{I_z}, \quad \text{and} \quad
\hat{H_1}=(A_\parallel+\omega_L) \hat{I_z} + B_\perp \hat{I_x}.
\end{equation}

All nuclear gates are implemented by applying the a sequence of
periodic pulses on the electron spin:
\begin{equation}
(\tau-\pi-2\tau-\pi-\tau)^{N/2},
\end{equation}
with $\tau$ a free evolution time, $\pi$ a pi-pulse on the
electron and $N$ the total number of pulses. We symmetrize the
decoupling sequence by alternating pi-pulses around the $X$ and
$Y$ axis (base sequence $X-Y-X-Y-Y-X-Y-X$, which is then
repeated). The nuclear evolution operators for the basic sequence
($N=2$) are:
\begin{eqnarray}
\label{eq:V01}
\hat{V_0}&=&\exp{[-i\hat{H_0} \tau]} \exp{[-i \hat{H_1} 2\tau]} \exp{[-i\hat{H_0} \tau]}\\
\hat{V_1}&=&\exp{[-i\hat{H_1} \tau]} \exp{[-i \hat{H_0} 2\tau]}
\exp{[-i\hat{H_1} \tau],}
\end{eqnarray}
for $m_s=0$ and $m_s=-1$ respectively.

The conditional operators $\hat{V_0}$ and $\hat{V_1}$ can be
represented as:
\begin{eqnarray}
\hat{V_0}&=&\exp{[-i\phi (\mathbf{\hat I} \cdot \mathbf{\hat{n}}_0)]}\\
\hat{V_1}&=&\exp{[-i\phi (\mathbf{\hat I} \cdot
\mathbf{\hat{n}}_1)],} \label{eq:V02}
\end{eqnarray}
which illustrates that the net evolution is a rotation by an angle
$\phi$ around an axis $\mathbf{\hat{n}}_i$ that depends on the
initial state of the electron spin: $\mathbf{\hat{n}}_0$ for
$m_s=0$ and $\mathbf{\hat{n}}_1$ for $m_s=-1$. The rotation angle
$\phi$ is independent of the electron spin input state
\cite{Taminiau2012}. Next, we show that both conditional and
unconditional rotations can be constructed by choosing $\tau$.

\begin{figure}[t]
\begin{center}
\includegraphics[scale=0.85]{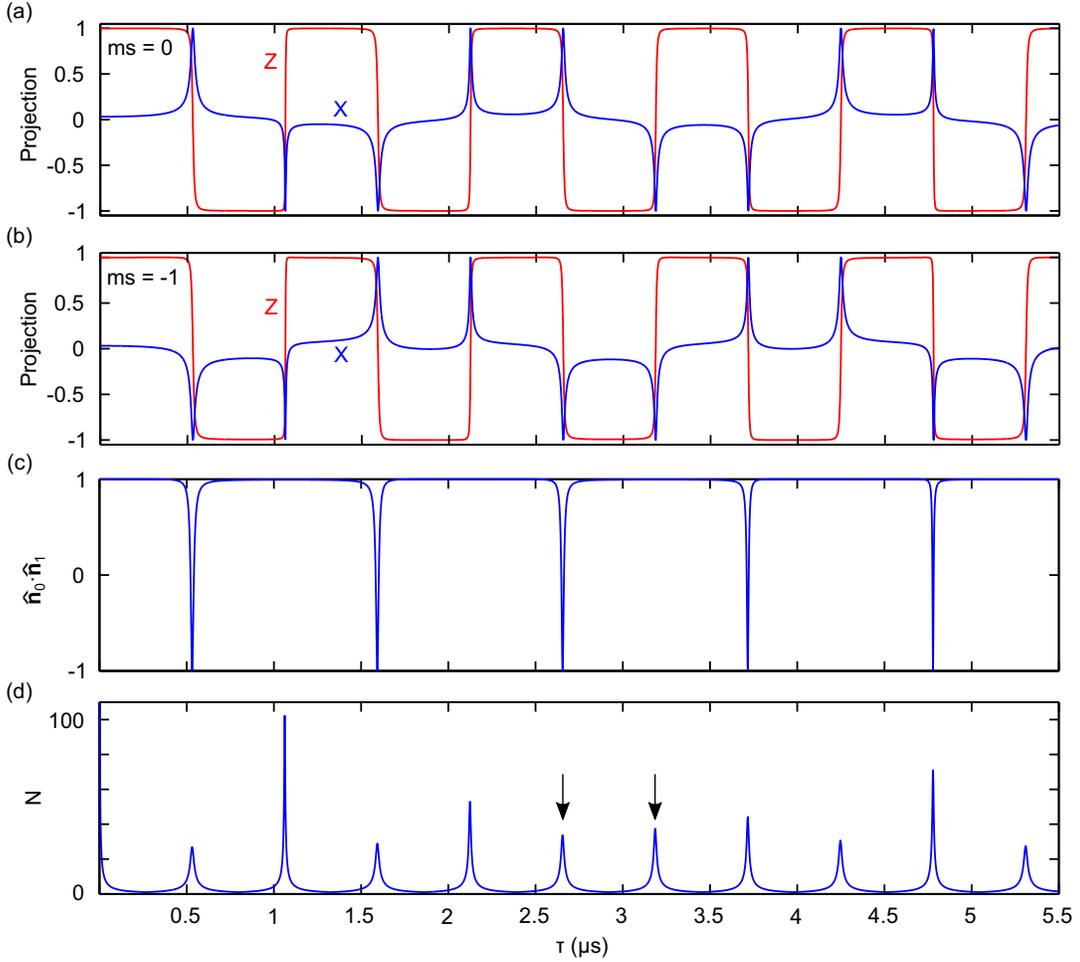}
\end{center}
\caption{\textbf{Simulations of the nuclear dynamics for spin 1.}
(a-b) The $X$ and $Z$ projections of (a) $\mathbf{\hat{n}}_0$ (the
net rotation axis for initial electron state $m_s=0$) and of (b)
$\mathbf{\hat{n}}_1$ (initial electron state $m_s=-1$). The $Y$
projection is 0. (c) The inner product $\mathbf{\hat{n}}_0 \cdot
\mathbf{\hat{n}}_1$ of the the two rotation axis indicates if the
gate is unconditional (parallel axes, $\mathbf{\hat{n}}_0 \cdot
\mathbf{\hat{n}}_1=1$) or conditional (anti-parallel axes,
$\mathbf{\hat{n}}_0 \cdot \mathbf{\hat{n}}_1=-1$). (d) The number
of pulses $N$ required for a $\pi/2$-rotation. The total gate
duration is given by $2N\tau$. The two arrows mark the values for
$\tau$ for the conditional and unconditional gates for this spin.}
\label{Fig:Simulatons_Carbon1} \vspace*{-0.2cm}
\end{figure}

\begin{figure}[t]
\begin{center}
\includegraphics[scale=0.85]{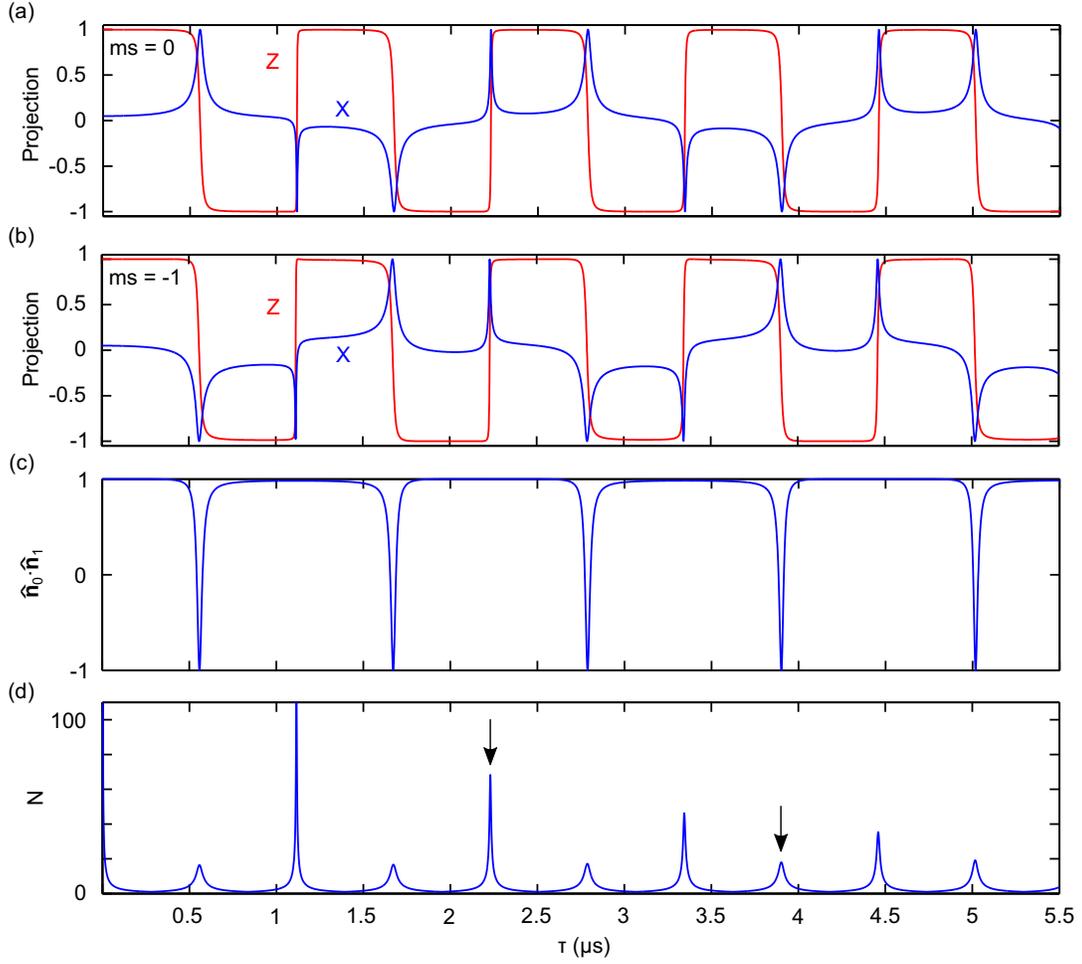}
\end{center}
\caption{\textbf{Simulations of the dynamics for nuclear spin 2.}
See description in figure \ref{Fig:Simulatons_Carbon1}.}
\label{Fig:Simulatons_Carbon3} \vspace*{-0.2cm}
\end{figure}

Figure \ref{Fig:Simulatons_Carbon1} shows the dynamics for nuclear
spin 1. Because $\omega_L >> A_\perp$, the $X$ and $Z$ components
of the rotation axes $\mathbf{\hat{n}}_0$ (Fig.
\ref{Fig:Simulatons_Carbon1}a) and $\mathbf{\hat{n}}_1$ (Fig.
\ref{Fig:Simulatons_Carbon1}b) show sharp resonances, for which
the nuclear spin undergoes an $X$-rotation. These resonances occur
for:
\begin{equation}
\tau \approx \frac{k\pi}{2\omega_L+A_\parallel},
\end{equation}
with integer $k$. The $X$-rotation is conditional for the odd
resonances (odd $k$, antiparallel rotation axes:
$\mathbf{\hat{n}}_0 \cdot \mathbf{\hat{n}}_1 = -1$) and
unconditional for the even resonances (even $k$, parallel axes:
$\mathbf{\hat{n}}_0 \cdot \mathbf{\hat{n}}_1 = 1$) (Fig.
\ref{Fig:Simulatons_Carbon1}c). For all other values of $\tau$ the
nuclear spin undergoes a simple $Z$-rotation independent of the
electron spin state ($\mathbf{\hat{n}}_0 \cdot \mathbf{\hat{n}}_1
= 1$). The electron and nuclear spin are then effectively
decoupled from each other. The number of pulses $N$ required for a
 $\pi/2$ rotation are shown in figure
\ref{Fig:Simulatons_Carbon1} as a function of $\tau$. The dynamics
for spin 2 are similar (Fig. \ref{Fig:Simulatons_Carbon3}), but
the resonances occur for different values of $\tau$ due to the
difference in $A_\parallel$.

The values for $\tau$ and $N$ for the gates used in this work are
given in Table \ref{Table:GateDesignParameters} and the values for
$\tau$ are also indicated in Figures \ref{Fig:Simulatons_Carbon1}d
and \ref{Fig:Simulatons_Carbon3}d.

The sharp resonances enable the universal control of a selected
nuclear spin, while decoupling the electron spin from all other
nuclear spin qubits and the rest of the environment. The gates are
thus selective, not limited by the electron $T_2^*$ or $T_2$ and
do not require strong coupling.

\begin{table}
  \centering
   \begin{tabular}{cc|c|c|c}
      \       &   \            & $\tau$ ($\mu$s)   & N             &   Total time ($\mu$s, rounded)   \\ [1mm]
      \hline
      Spin1:  & $R_X^e(\pi/2)$ & $\ \ $2.656$\ \ $ & $\ \ $32$\ \ $ & $\ \ $170$\ \ $ \\
      \       & $R_X(\pi/2)$   & $\ \ $3.186$\ \ $ & $\ \ $40$\ \ $ & $\ \ $255$\ \ $ \\
      \       & $R_Z(\pi/2)$   & $\ \ $2.058$\ \ $ & $\ \ $4$\ \ $  & $\ \ $16$\ \ $\\
      \hline
      Spin2:  & $R_X^e(\pi/2)$ & $\ \ $3.900$\ \ $ & $\ \ $18$\ \ $ & $\ \ $140$\ \ $\\
      \       & $R_X(\pi/2)$   & $\ \ $2.228$\ \ $ & $\ \ $64$\ \ $ & $\ \ $285$\ \ $ \\
      \       & $R_Z(\pi/2)$   & $\ \ $2.100$\ \ $ & $\ \ $2$\ \ $  & $\ \ $8$\ \ $\\
  \end{tabular}
  \caption{\textbf{Gate parameters.} $R_\alpha(\theta)$ is a
  rotation of the nuclear spin around Bloch-sphere axis $\alpha$ by an angle $\theta$. For gates marked $R^e$ the rotation direction is controlled
  by the electron spin state, for all other gates the direction is unconditional.} \label{Table:GateDesignParameters}
\end{table}

\subsection{Nuclear gate characterization}

\begin{figure}[t]
\begin{center}
\includegraphics[scale=0.85]{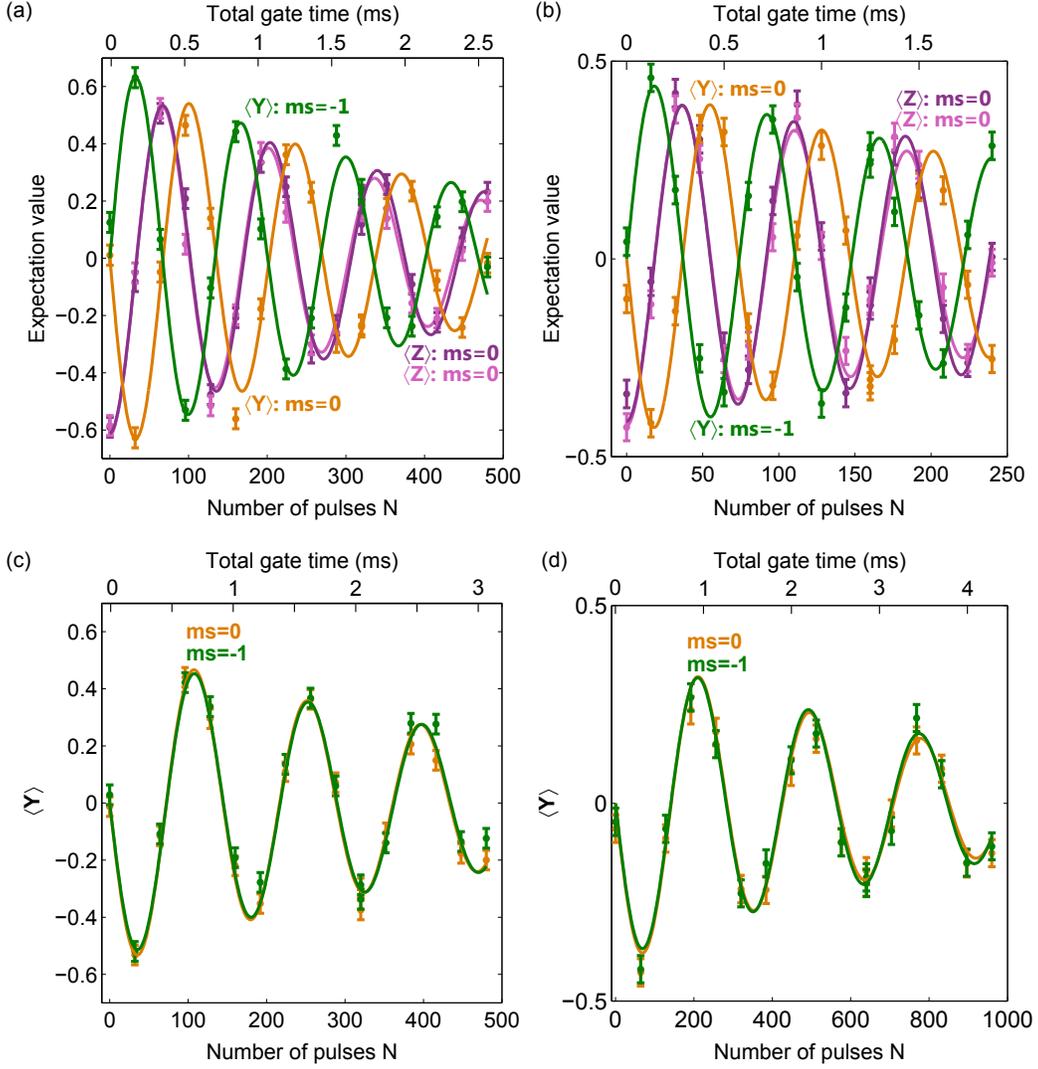}
\end{center}
\caption{\textbf{Gate characterization.} Action of the conditional
(a-b) and unconditional (c-d) gates for both nuclear spin 1 (a-c)
and nuclear spin 2 (b-d). Measurement schemes in figure 2e and 2f
of the main text. The $Z$-projection shows that the spin undergoes
a Rabi oscillation and the $Y$-projection shows that the
orientation of the rotation is either conditional (a-b) or
unconditional on the electron state (c-d). Not corrected for
initialization or readout fidelities.} \label{Fig:Gate
characterization} \vspace*{-0.2cm}
\end{figure}

To characterize the conditional and unconditional gates we study
the effect of the gates on an initialized nuclear spin state, as a
function of the number of electron spin pulses in the gate.
Figures 2e and 2f of the main text give the $Y$-projections for
both gates for nuclear spin 1. Figure \ref{Fig:Gate
characterization} gives the complete set of measurements,
including the gates for nuclear spin 2 and the $Z$-projections
that confirm that the gates are conditional and unconditional
rotations around $X$.

\subsection{Theoretical gate fidelities}

We calculate the theoretical fidelities for the action of the
gates on input state $\ket{0}$ (Table
\ref{Table:Fidelities_Spin1}). The results are seperated in 3
parts. The first column shows the theoretical fidelity directly as
obtained from $\hat{V_0}$ and $\hat{V_1}$. The second column takes
into account that the third nuclear spin has a parallel component
of the hyperfine interaction $A_\parallel$ that differs by less
than $10$ kHz from nuclear spin 2. This affects the conditional
rotation for nuclear spin 2, because the electron spin also
entangles slightly with nuclear spin 3. The third column
additionally includes the effect of the discretization of $\tau$
(experimental resolution of 2 ns). Note that the $Z$ gates are not
significantly affected by this because they do not rely on sharp
resonances. These fidelities do not take into account the rest of
the spin bath or phonon-induced decoherence or depolarization.

\begin{table}
  \centering
   \begin{tabular}{cc|c|c|c|c}
       \            &      \        &  Fidelity   & + Spin 3   & + Discretization precision \\[1mm]
      \hline
       Spin 1:      &$R_X^e(\pm\pi/2)$ &   0.996          &            &     0.987   \\
       \            &$R_X(\pi/2)$   &   0.997          &            &     0.993   \\
       \            &$R^Z(\pi/2)$   &   0.999          &            &     0.999   \\
      \hline
       Spin 2:      &$R_X^e(\pm\pi/2)$ &   0.997          &    0.959   &     0.953   \\
       \            &$R_Z(\pi/2)$   &   0.999          &            &     0.999   \\
  \end{tabular}
  \caption{\textbf{Theoretical fidelities for the gates.} The state fidelity with the target state after applying the gate
  on $\ket{0}$ and tracing out the electron state. Conditional $X$-rotation: $R_X^e(\pm\pi/2)$.
  Unconditional $X$- and $Z$-rotations: $R_X(\pi/2)$ and $R_X(\pi/2)$.}\label{Table:Fidelities_Spin1}
\end{table} 

\section{Nuclear initialization fidelity}

\begin{figure}[t]
\begin{center}
\includegraphics[scale=0.75]{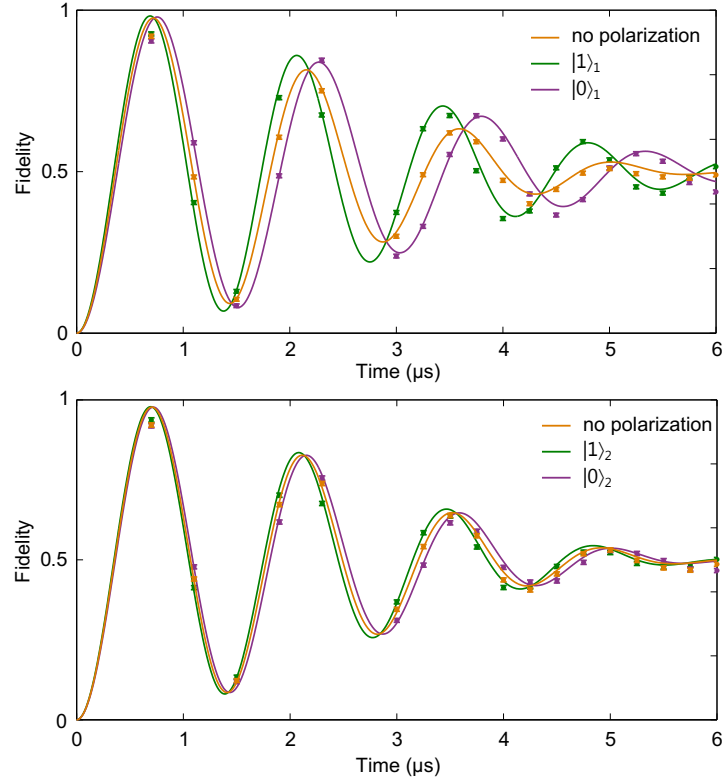}
\end{center}
\caption{\textbf{Data set for the determination of the nuclear
spin initialization fidelity.} Electron spin free evolution
measurements with and without nuclear spin initialization. Top:
nuclear spin 1, fidelity $F_1 = 0.91(2)$. Bottom nuclear spin 2,
fidelity $F_2 = 0.88(5)$. The three curves are measured in a
single experiment that is long enough to average over the magnetic
field fluctuations, reducing $T_2^*$ to 3.1(1) $\mu$s.}
\label{Fig:eRamsey_added} \vspace*{-0.2cm}
\end{figure}

The nuclear initialization fidelity is determined from Ramsey-type
experiments as described in the main text. The measurements in
Figure 1f and g of the main text are post selected on small
magnetic field drifts so that the absolute increase of $T_2^*$ can
be determined. For the initialization fidelity we use an average
over a larger data set (Fig. \ref{Fig:eRamsey_added}). This
approach has the advantage that the initialization fidelity can be
more accurately determined, but is not suited for measurements of
the absolute increase of $T_2^*$, due to significant magnetic
field fluctuations over the extended measurement time. We find
$F_1 = 0.91(2)$ for nuclear spin 1 and $F_2 = 0.88(5)$ for nuclear
spin 2.

\begin{figure}[t]
\begin{center}
\includegraphics[scale=0.85]{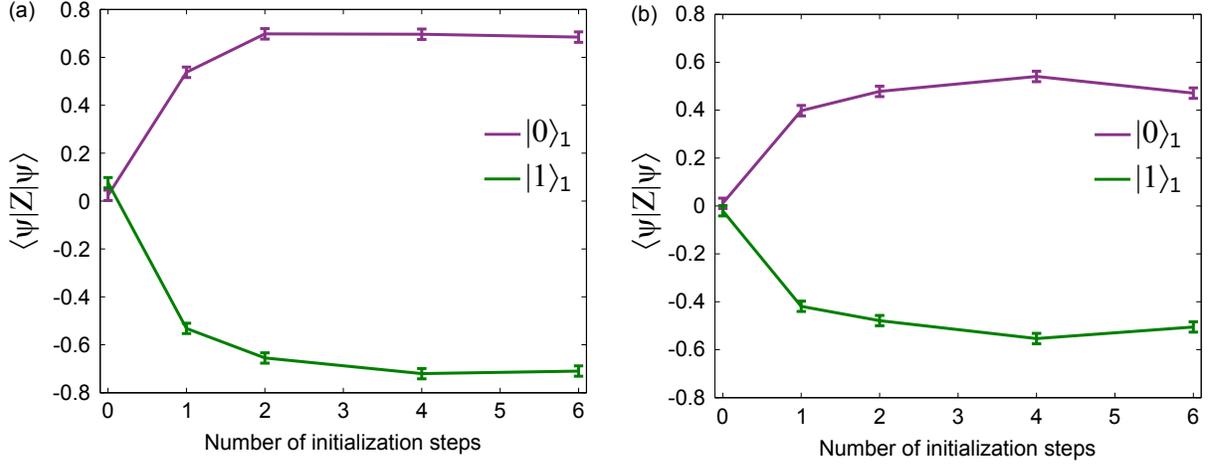}
\end{center}
\caption{\textbf{Increased initialization fidelity through
multiple initialization steps.} The measured expectation value
$\langle{Z}\rangle = \langle \psi |Z| \psi \rangle$ for the
nuclear spin state $\ket{\psi}$ as a function of the number of
initialization steps. The nuclear spin is either initialized on
$\ket{0}$ or $\ket{1}$. (a) Nuclear spin 1. (b) Nuclear spin 2.
Data not corrected for initialization or readout fidelities.}
\label{Fig:Nuclear_Initialization} \vspace*{-0.2cm}
\end{figure}

Although the initialization protocol ideally needs only a single
application, figure \ref{Fig:Nuclear_Initialization} shows that
repeated applications do further increase the polarization before
saturating after approximately 2 steps. In the implementation of
the quantum error correction protocol (Fig. 4 of the main text)
only a single initialization step was used so that the
initialization fidelities are lower than those obtained from
figure \ref{Fig:eRamsey_added}. The results in figure
\ref{Fig:Nuclear_Initialization} yield an initialization fidelity
for these experiments of $F_1 \approx F_2 \approx 0.82$.

\section{Two-qubit tomography}

\begin{figure}[t]
\begin{center}
\includegraphics[scale=0.6]{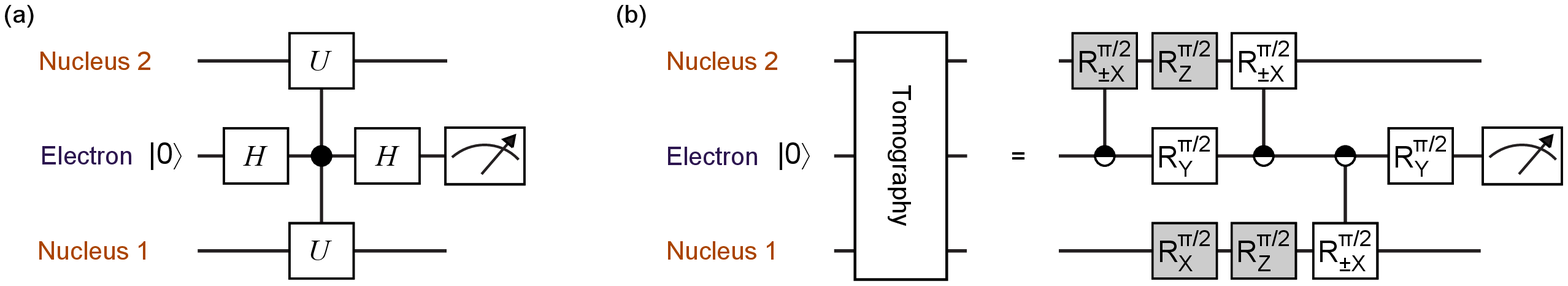}
\end{center}
\caption{\textbf{Two-qubit tomography.} We measure the expectation
values of the different combinations of the nuclear two-qubit
pauli operators using the electron spin. (a) General readout
scheme to measure $\langle UU\rangle$, with $U$ a unitary operator
and $H$ the Hadamard gate. (b) Our implementation. The shaded
gates are optional basis rotations.} \label{Fig:2QubitTomo}
\vspace*{-0.2cm}
\end{figure}

Two-qubit tomography (main text figure 3) is performed by mapping
two-qubit correlations onto the electron spin before reading out
the electron. Figure \ref{Fig:2QubitTomo}a shows the general
principle and figure \ref{Fig:2QubitTomo}b shows our
implementation.

\section{Quantum error correction}

This section discusses the application and characterization of the
errors, gives the derivation of the theoretical analysis used in
the main text, and gives the complete set of state fidelity
results used to derive the process fidelities in the main text
(main text Figure 4).

\subsection{Error implementation}

\begin{figure}[t]
  \includegraphics[scale=0.6]{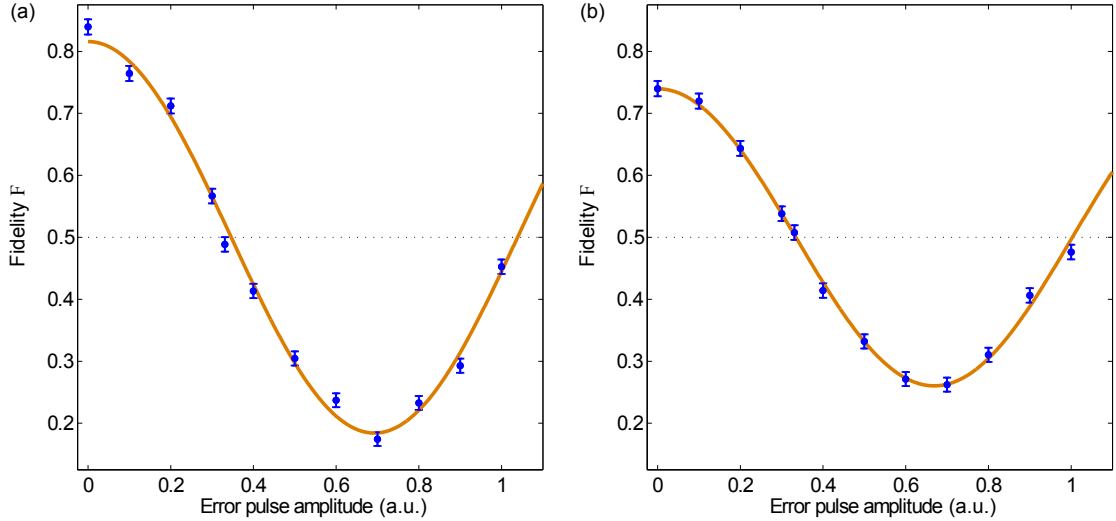}\\
  \caption{\textbf{Characterization of the nuclear spin errors}.
  Fidelity $F$ of the final nuclear state with $\ket{Z} = \ket{0}$ as a function of the amplitude of the error pulse applied on the electron.
  (a) For nuclear spin 1 and (b) for nuclear 2. Lines are fits to the
  expected cosine behavior around fidelity 0.5. The measurements are not
  corrected for readout or initialization
fidelities.}\label{Nuclear_Spin_Error_Calibration}
\end{figure}

The quantum error correction scheme corrects both coherent errors
of the type $R_X(\theta)$, i.e. a rotation around $X$ by angle
$\theta$ or the quantum map:

\begin{equation}\label{Error_Quantum_Map}
\epsilon_c(\rho, \theta) = [\cos(\theta/2)I + i\sin(\theta/2)X]
\rho [\cos(\theta/2)I - i\sin(\theta/2)X]
\end{equation}

and non-unitary, decoherence-type, operations given by the quantum
map:

\begin{equation}\label{Error_Quantum_Map_Incoherent}
\epsilon_d(\rho,\theta) = \cos^2(\theta/2)I\rho{I} +
\sin^2(\theta/2)X\rho{X}
\end{equation}

Errors on the electron spin are directly created by a microwave
pulse that implements $R_X(\theta)$. In half the experiments a
positive rotation $R_X(+\theta)$ is applied and in the other half
a negative rotation $R_X(-\theta)$. The final result is the
average over the two measurements, so that the implemented error
is of the form of equation \ref{Error_Quantum_Map_Incoherent}.

Errors on the nuclear spins are implemented through the electron
spin. First the error is applied to the electron spin. Then the
initialization gate transfers the error to the nuclear spin state.
This operation gives the same result as applying $\epsilon_d(\rho,
\theta)$ on an initialized nuclear spin state. Figure
\ref{Nuclear_Spin_Error_Calibration} shows the resulting fidelity
with $\ket{Z}=\ket{0}$ for both nuclear spins as a function of the
error pulse amplitude. The observed oscillation around $F=1/2$
confirms the expected application of the error.

\subsection{Theoretical analysis}

We analyze the quantum error process as a combination of imperfect
majority voting and a general decoherence/depolarization process.
We assume the following two properties: (1) that applied errors
have no effect on the $\ket{\pm X}$ states and (2) that the
probabilities that the error correction (majority voting) is
successful if qubit 1, qubit 2 or qubit 3 is different from the
other two qubits are given by the three $p_1$, $p_2$ and $p_3$
respectively. These probabilities then completely describe the
effectiveness of the error correction process (for an ideal case
$p_1 = p_2 = p_3 = 1$).

In each experiment we prepare 6 input states $\ket{\alpha}$:\\

\begin{align}
\label{eq:states}
\ket{Z} = \ket{0},\\
\ket{-Z} = \ket{1},\\
\ket{X} = \frac{1}{\sqrt{2}}(\ket{0}+\ket{1}),\\
\ket{-X} = \frac{1}{\sqrt{2}}(\ket{0}-\ket{1}),\\
\ket{Y} = \frac{1}{\sqrt{2}}(\ket{0}+i\ket{1}),\\
\ket{-Y} = \frac{1}{\sqrt{2}}(\ket{0}-i\ket{1}),
\end{align}
and measure the expectation values:
\begin{equation}
C_\alpha = \langle{\psi_\alpha}|\alpha|{\psi_\alpha}\rangle,
\end{equation}
where $\ket{\psi_\alpha}$ is the output state for input state
$\ket{\alpha}$, and $\alpha = Z,\ -Z,\ Y,\ -Y,\ X,$ or $-X$. The
fidelities of the output states with the input states are given
by:
\begin{equation}
F_\alpha = C_\alpha/2 +1/2.
\end{equation}

We label the 8 possible combinations of (applied) errors that can
occur with $j$. For example: $j=000$ implies no error, $j=100$ is
an error on Qubit 1, etc. The obtained signal for error
combination $j$ and input state $\ket{\alpha}$ is $C_\alpha^j$.
Using the above assumptions all possible results can be described
by the probabilities $p_1$,$p_2$ and $p_3$ together with the
obtained signals when no error is applied $C_{\pm Z}^{000}$,
$C_{\pm Y}^{000}$ and $C_{\pm X}^{000}$. All values based on the
above error correction model are given in Table \ref{Table:8
Signals}.

\begin{table}[h]
  \centering
   \begin{tabular}{c|c|c|c}
      Signal for error $j$  &  $\ket{\pm Z}$ & $\ket{\pm Y}$ & $\ket{\pm X}$ \\[1mm]
      \hline
            & & & \\
      $C_\alpha^{000}$ & $\ \  C_{\pm Z}^{000} \          \ $ & $\ \  C_{\pm Y}^{000} \ \            $ & $\ \  C_{\pm X}^{000} \ \ $\\[1mm]
      $C_\alpha^{001}$ & $\ \  (2p_3-1)C_{\pm Z}^{000} \  \ $ & $\ \  (2p_3-1)C_{\pm Y}^{000} \ \    $ & $\ \  C_{\pm X}^{000} \ \ $\\[1mm]
      $C_\alpha^{010}$ & $\ \  (2p_2-1)C_{\pm Z}^{000} \  \ $ & $\ \  (2p_2-1)C_{\pm Y}^{000} \ \    $ & $\ \  C_{\pm X}^{000} \ \ $\\[1mm]
      $C_\alpha^{100}$ & $\ \  (2p_1-1)C_{\pm Z}^{000} \  \ $ & $\ \  (2p_1-1)C_{\pm Y}^{000} \ \    $ & $\ \  C_{\pm X}^{000} \ \ $\\[1mm]
      $C_\alpha^{011}$ & $\ \  -(2p_1-1)C_{\pm Z}^{000} \ \ $ & $\ \  -(2p_1-1)C_{\pm Y}^{000} \ \   $ & $\ \  C_{\pm X}^{000} \ \ $\\[1mm]
      $C_\alpha^{101}$ & $\ \  -(2p_2-1)C_{\pm Z}^{000} \ \ $ & $\ \  -(2p_2-1)C_{\pm Y}^{000} \ \   $ & $\ \  C_{\pm X}^{000} \ \ $\\[1mm]
      $C_\alpha^{110}$ & $\ \  -(2p_3-1)C_{\pm Z}^{000} \ \ $ & $\ \  -(2p_3-1)C_{\pm Y}^{000} \ \   $ & $\ \  C_{\pm X}^{000} \ \ $\\[1mm]
      $C_\alpha^{111}$ & $\ \  -C_{\pm Z}^{000} \         \ $ & $\ \  -C_{\pm Y}^{000} \ \           $ & $\ \  C_{\pm X}^{000} \ \ $\\
   \end{tabular}
  \caption{Action of the error correction protocol. $C_\alpha^{j}$ is the signal obtained
  for input state $\ket{\alpha}$ and error combination $j$. $p_n$ is the probability that
  an error on qubit $n$ is successfully corrected.}\label{Table:8 Signals}
\end{table}

\subsubsection*{Single-qubit errors}

For a variable strength error on one of the qubits the final
fidelity for inputs $\ket{\pm Y}$ and $\ket{\pm Z}$ is given by a
weighted sum of the two corresponding values in table \ref{Table:8
Signals}:

\begin{equation}\label{Fit_Equation1}
F_\alpha(\theta) = \frac{\cos^2(\theta/2)}{2}C_\alpha^{klm} +
\frac{\sin^2(\theta/2)}{2}C_\alpha^{k'l'm'} + 1/2,
\end{equation}

in which $klm$ and $k'l'm'$ identify the applied error
combination. For example, for the variable error applied to qubit
2 and no error to qubits 1 and 3, we have $klm=000$ and $k'l'm' =
010$. For $\ket{\pm X}$ the signal is simply constant:

\begin{equation}\label{Fit_Equation2}
   F_{\pm X}(\theta) = C_{\pm X}^{000}/2 + 1/2.
\end{equation}

In figure 4d of the main text two different types of errors are
applied: (1) just a variable error on qubit $n$ and (2) a variable
error on qubit 2 and a full flip on qubit 1. For a variable error
on Qubit $n$ ($n=1,2,3$) and input $\ket{\pm Y}$ or $\ket{\pm Z}$
equation \ref{Fit_Equation1} simplifies to:

\begin{equation}\label{Fit_Equation3}
   F_{\alpha}(\theta) = \frac{C_\alpha^{000}}{2}(p_n + (1-p_n)\cos(\theta)) +
   1/2,
\end{equation}

in agreement with the interpretation of the values $p_n$ as the
probability that an error on qubit $n$ is successfully corrected.
For the variable error on qubit 2 and a full flip on qubit 1 we
find:

\begin{equation}\label{Fit_Equation4}
   F_{\alpha}(\theta) = \frac{C_\alpha^{000}}{2}(p_1 - p_3  + (p_1+p_3-1)\cos(\theta)) +
   1/2,
\end{equation}

which is of the same form as equation \ref{Fit_Equation3} and
shows that for $p_1 = p_3$ a cosine around fidelity $1/2$ is
obtained; the error correction is effectively switched of.

The process fidelity $F_p$ of the error correction process with
the identity is:

\begin{equation}\label{Eq:ProcessFidelity}
F_{p}(\theta)=\frac{F_Z(\theta) + F_{-Z}(\theta) + F_X(\theta) +
F_{-X}(\theta) + F_Y(\theta) + F_{-Y}(\theta)}{4} - 1/2,
\end{equation}

in which the $F_\alpha$ are given by equations \ref{Fit_Equation1}
and \ref{Fit_Equation2}. For a single applied error this
simplifies to:

\begin{equation}\label{Eq:ProcessFidelity_This_case}
F_{p}(\theta)= F_{p0} + A_{YZ}(p_n + (1-p_n)\cos(\theta)),
\end{equation}
in which $F_{p0} = (F_X + F_{-X})/4$ and $A_{YZ} = (F_Y(0) +
F_{-Y}(0) + F_Z(0) + F_{-Z}(0) -2)/4$. Note that all the different
fidelities without error get grouped into two constants, one
related to the average fidelity of the $\ket{\pm X}$ states and
one related to the average fidelity of the $\ket{\pm Y}$ and
$\ket{\pm Z}$ states without applied errors.

\subsubsection*{Ancilla initialization fidelity}

\begin{figure}[t]
  \includegraphics[scale=0.6]{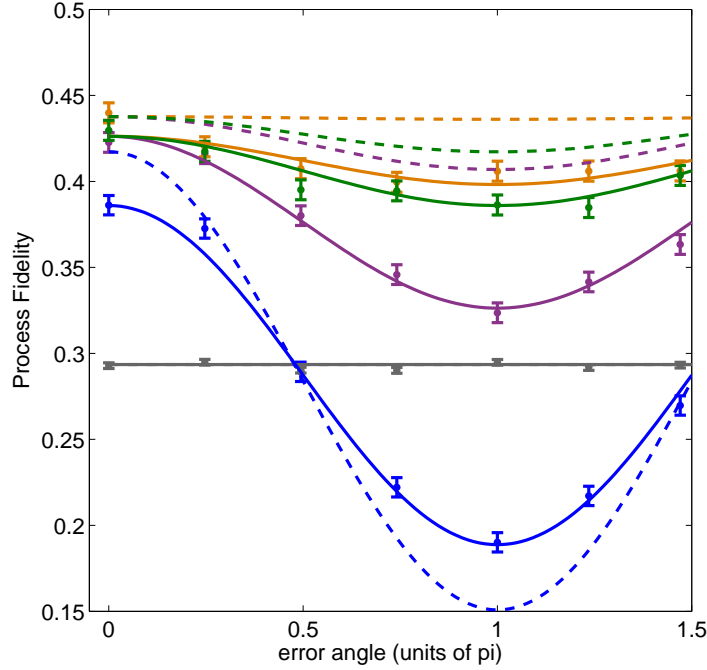}\\
  \caption{\textbf{Effect of initialization on the error correction}. Process
  fidelity for selectively applied errors with error angle $\theta$ as in figure 4d of the main text.
  Solid lines are a fit to the model including equation \ref{Initialization_Matrix} .
  The dashed lines are the expected results
  for ideal initialization of the two nuclear
  ancillas ($F_1=F_2=1$).}\label{Selected_Errors_Process_Corrected}
\end{figure}

The above model can be modified to explicitly take the effect of
imperfect initialization/polarization of the ancilla qubits (qubit
1 and 3) into account. The effect of the initialization fidelities
$F_{1}$ and $F_{2}$ of the two ancilla qubits is that the measured
values $C'_\alpha$ are now combinations of the $C_\alpha$ values
in Table \ref{Table:8 Signals} following:

\begin{equation}
\begin{bmatrix} C_\alpha^{'0k0} \\ C_\alpha^{'0k1} \\ C_\alpha^{'1k0} \\ C_\alpha^{'1k1}
\end{bmatrix} =
\begin{bmatrix} F_{1}F_{2} & F_{1}(1-F_{2})& (1-F_{1})F_{2} & (1-F_{1})(1-F_{2}) \\
F_{1}(1-F_{2})& F_{1}F_{2} & (1-F_{1})(1-F_{2}) & (1-F_{1})F_{2}
\\  (1-F_{1})F_{2} & (1-F_{1})(1-F_{2}) & F_{1}F_{2} &
F_{1}(1-F_{2}) \\ (1-F_{1})(1-F_{2}) & (1-F_{1})F_{2} &
F_{1}(1-F_{2}) & F_{1}F_{2} \end{bmatrix}
\begin{bmatrix} C_\alpha^{0k0}  \\ C_\alpha^{0k1} \\ C_\alpha^{1k0} \\ C_\alpha^{1k1}
\end{bmatrix}\label{Initialization_Matrix}
\end{equation}

This extended model separates the initialization imperfections
from imperfections in the error correction process. We take $F_1 =
F_2 = 0.82$ as an estimate for the initialization fidelities (Fig.
\ref{Fig:Nuclear_Initialization}). The resulting fits are shown in
figure \ref{Selected_Errors_Process_Corrected} (solid lines) and
yield $p_n = 0.93(3), 0.89(3),0.99(3)$ and $\langle{p_n}\rangle =
(p_1+p_2+p_3)/3 = 0.94(2)$. We calculate the expected result for
ideal initialization by using the same value for $p$ but now
setting $F_1=F_2=1$ (dashed lines, Fig.
\ref{Selected_Errors_Process_Corrected}). The imperfect
initialization has two effects. First it strongly affects the
success probability of the error correction as double errors (one
initialization error + one applied error) cannot be corrected.
Second, it lowers the overall maximum fidelity. This is a weak
effect because it requires an error in the preparation of both
ancillas at the same time and is therefore proportional to
$(1-F_1)(1-F_2)$.

\begin{figure}[t]
  \includegraphics[scale=0.6]{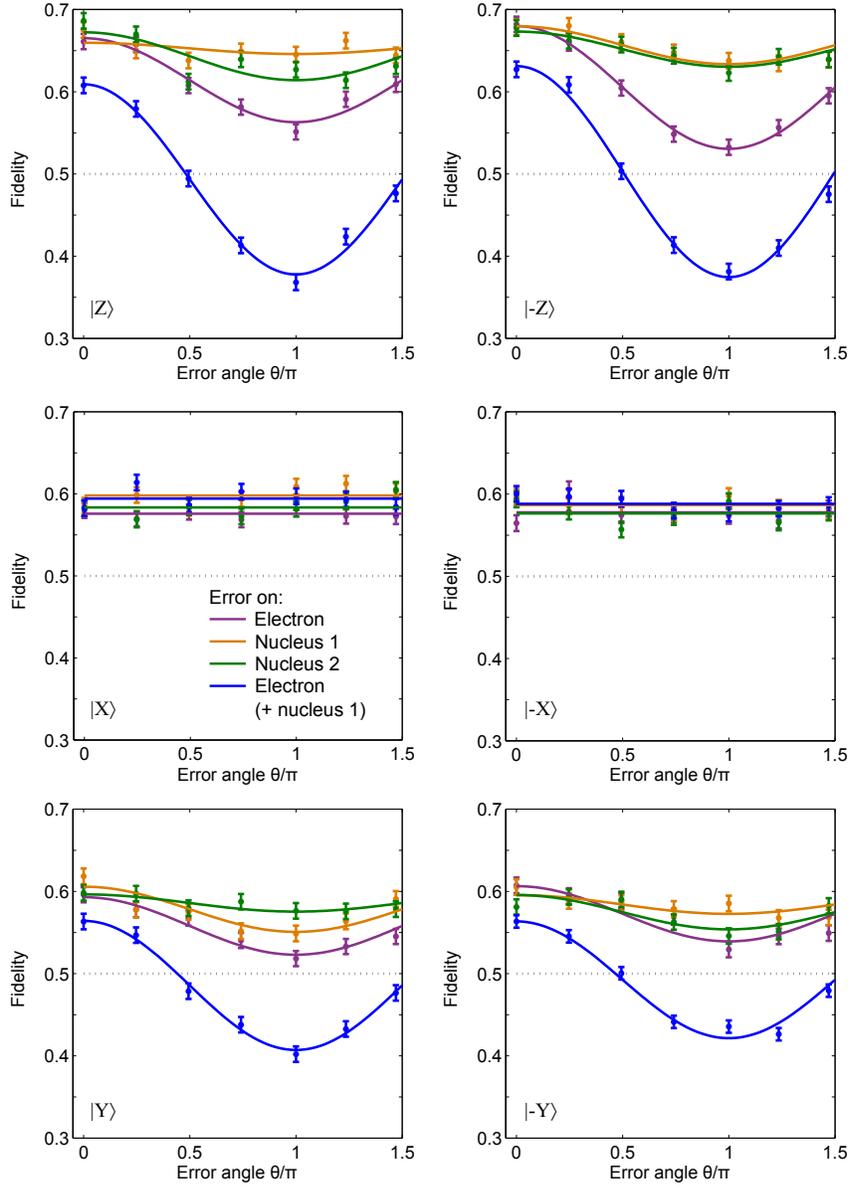}\\
  \caption{\textbf{Complete set of state fidelities for selectively applied errors.} State fidelity
  for the $6$ input states $\ket{Z}, \ket{-Z}, \ket{X}, \ket{-X}, \ket{Y}$ and
  $\ket{-Y}$, for $4$ different combinations of errors and as a function of the error angle $\theta$.
  We apply $\epsilon_d(\rho,\theta)$ to each of the three qubits
  separately and a combination of $\epsilon_d(\rho,\theta)$ to the
  electron and $\epsilon_d(\rho,\pi)$ to Nucleus 1. Lines are fits to equations \ref{Fit_Equation2}-\ref{Fit_Equation4}.
  }\label{Selected_Errors}
\end{figure}

\newpage

\subsubsection*{Simultaneous errors}

The process fidelity $F_p$ for simultaneous errors is given by:

\begin{align}\label{Process_Fidelity_Simul1}
F_{p}(p_e)  =& (1-3p_e+3p_e^2-p_{e}^3)F_p^{000} \\
&+p_{e}(1-p_{e})^2(F_p^{001}+F_p^{010}+F_p^{100}) \\
&+p_{e}^2(1-p_{e})(F_p^{011}+F_p^{101}+F_p^{110})\\
&+p_{e}^3F_p^{111},
\end{align}

with $p_e = \sin(\theta/2)^2$ the error probability and
$F_p^{klm}$ the process fidelity for applied error $klm$, i.e.:

\begin{equation}\label{Process_Fidelity_Simul2}
F_p^{klm} = 1/4\left( 1 + \frac{C_X^{klm} + C_{-X}^{klm}}{2} +
\frac{C_Y^{klm} + C_{-Y}^{klm}}{2} + \frac{C_Z^{klm} +
C_{-Z}^{klm}}{2}\right).
\end{equation}

with $C_\alpha^{klm}$ as given in Table \ref{Table:8 Signals} we
obtain:

\begin{equation}\label{Process_Fidelity_Simul3}
F_{p}(p_e)  = F_{p0} +
A_{YZ}\left[1-3p_e+3p_e^2-2p_e^3+3(2\langle{p_n}\rangle-1)(p_e-3p_e^2+2p_e^3)\right],
\end{equation}

The experiment is completely described by just 3 parameters: the
offset $F_{p0}$ (due to the average $\ket{\pm X}$ fidelity without
applied errors), the amplitude $A_{YZ}$ (due to the average
$\ket{\pm Y,Z}$ fidelity without applied errors) and the average
error correction probability $\langle{p_n}\rangle$.

\subsection{Complete state fidelity data set}

\begin{figure}[t]
  \includegraphics[scale=0.6]{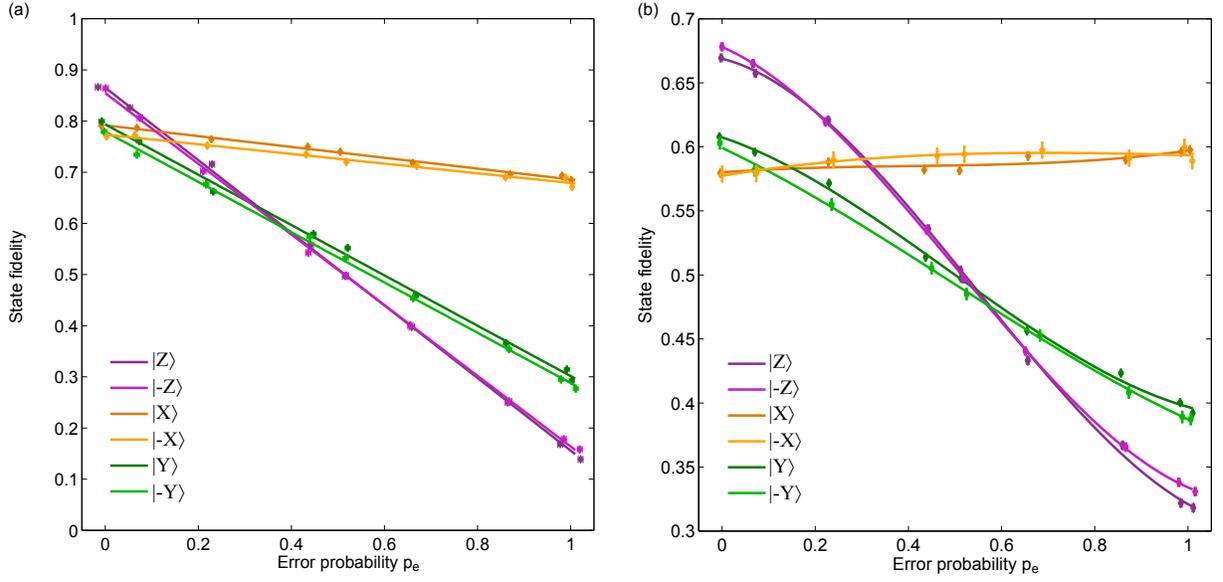}\\
  \caption{(a) State fidelities for the 6 input states as a function of the error probability
  without error correction. Lines are linear fits to the data. (b) State fidelities for the 6 input states as a function of the error probability
  with error correction. Lines are $3^{rd}$ order polynomial fits to the data. }\label{6_curves_ENCDEC}
\end{figure}

The complete set of state fidelities used to derive the process
fidelities for errors applied to one of the qubits at a time (main
text figure 4d) is shown in figure \ref{Selected_Errors}. The
complete sets of state fidelities used to obtain the process
fidelities for simultaneously applied errors (Fig. 4e of the main
text) are given in figure \ref{6_curves_ENCDEC}a (without error
correction) and figure \ref{6_curves_ENCDEC}b (with error
correction).

\section{Decoherence and depolarization}

In this section we analyze the different decoherence mechanisms in
the three-qubit register.

\subsection{Electron depolarization ``$T_1$''}

\begin{figure}[t]
\begin{center}
\includegraphics[scale=0.75]{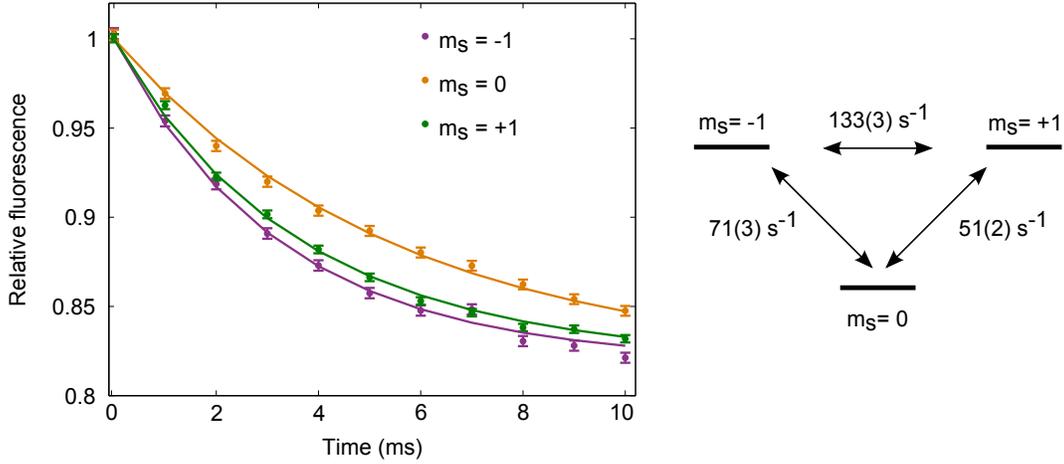}
\end{center}
\caption{\textbf{Electron depolarization data and fit to
three-level model.} The fluorescence of the final electron readout
is shown relative to the fluorescence for $m_s=0$ preparation. The
fits additionally use the measured fluorescence for $m_s=-1$ and
$m_s=+1$ preparation.} \label{Fig:T1_3level} \vspace*{-0.2cm}
\end{figure}

The electronic depolarization (longitudinal relaxation or T1-type
process) due to phonon interactions plays an important role at
these room temperature experiments. To measure the depolarization
rates we prepare one of the three states $m_s=-1$, $m_s=0$ and
$m_s=+1$ and let the system relax for a time $t$. We then apply a
pi-pulse on the $m_s=0$ transition (for the state starting in
$m_s=0$ nothing is done) before reading out the electron. The
results are fit to a 3-level model that yields three rates between
the different levels (Fig. \ref{Fig:T1_3level}). We find:
$\Gamma_{0,-1}=71(3)$ s$^{-1}$, $\Gamma_{0,+1}=51(2)$ s$^{-1}$ and
$\Gamma_{-1,+1}=133(3)$ s$^{-1}$. In this three level system no
unique ``$T_1$'' value can be defined. Nevertheless, a separate
analysis of each of the curves gives $1/e$ times of $3.24(9)$ ms
($m_s=-1$), $5.11(7)$ ms ($m_s=0$, which is often reported as the
$T_1$ value) and $3.91(6)$ ms ($m_s=+1$). We verified that the
same rates were obtained with a 4 times lower laser output power,
indicating that transitions induced by background illumination are
negligible.

\subsection{Electron decoherence $T_{coh}$}

\begin{figure}[t]
\begin{center}
\includegraphics[scale=0.6]{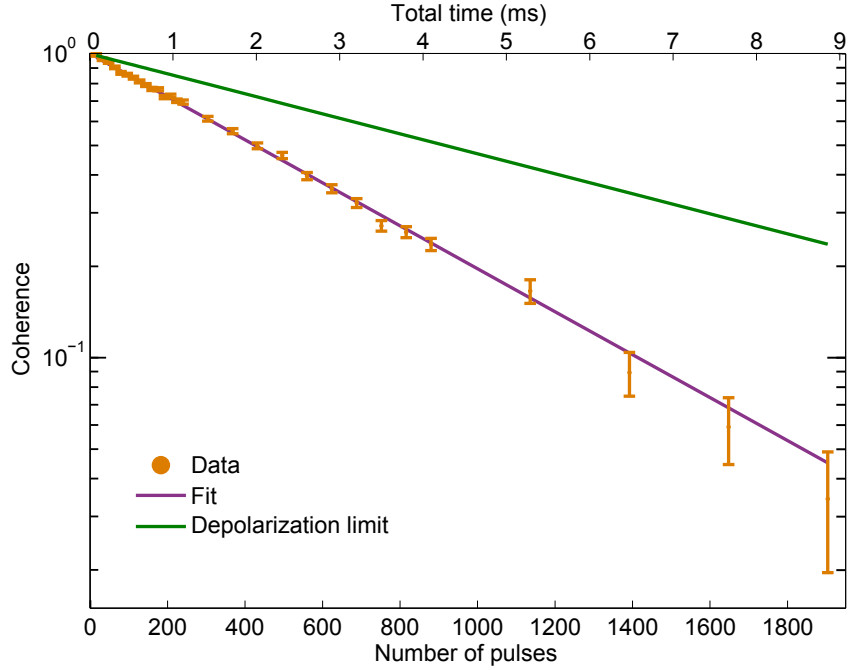}
\end{center}
\caption{\textbf{Decoherence of the electron spin under dynamical
decoupling.} We apply a decoupling sequence to input state
$\ket{X}$ and measure the final state along $X$. The interpulse
delay $\tau$ is $2\pi/\omega_L$, in the same range as used for the
nuclear gates. Purple: Exponential fit to the data that gives
$T_{coh}=2.86(4)$. Green: expected value due to electron
depolarization alone.} \label{Fig:DD_superposition}
\vspace*{-0.2cm}
\end{figure}

To measure the electronic coherence time under dynamical
decoupling $T_{coh}$ the electron spin is prepared along $X$. We
then apply a decoupling sequence with $\tau = 2\pi/\omega_L =
2.324\ \mu$s and measure the spin projection along $X$. The total
time is varied by varying the number of pulses $N$ in the
sequence. The result is shown in figure
\ref{Fig:DD_superposition}.

The green line marks the limit given by phonon-induced
depolarization of the electron spin. It is given by the total
decay rate out of the $m_s=0$ and $m_s=-1$ levels: $\Gamma_{0,-1}
+ \Gamma_{0,+1}/2 + \Gamma_{-1,+1}/2$. The additional decoherence
observed experimentally is consistent with previous reports
\cite{BarGill2012} and is likely due to phonon-induced dephasing,
as much longer coherence times were reported at low temperatures
\cite{Bernien2013,BarGill2012}. The expected signal without
phonon-induced dephasing and depolarization is given by:

\begin{equation}
S = e^{-\left(\frac{2\tau}{T_2}\right)^n N,
}\label{eq:eRamsey_naive}
\end{equation}

with the spin echo time $T_2 = 251(7) \mu$s. With $n=3$ this gives
an estimated decay time of $\sim 700$ ms, indicating that
decoupling from the spin bath is not the limiting factor.

\subsection{Nuclear dephasing $T_2^{\star}$}

\begin{figure}[t]
\begin{center}
\includegraphics[scale=0.7]{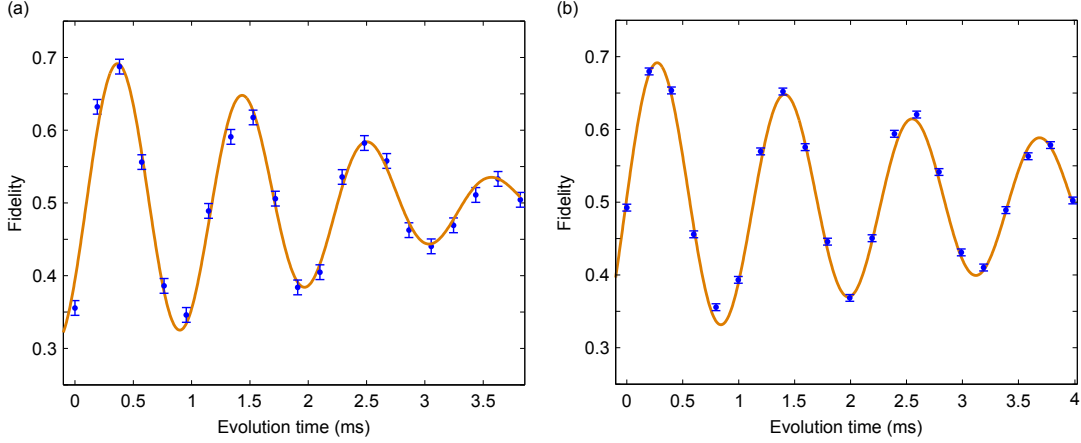}
\end{center}
\caption{\textbf{Nuclear dephasing time $T_2^\star$.} (a) For
nuclear spin 1. $T_2^{\star} = 2.7(2)$ ms, simultaneously measured
electron $T_2^{\star} = 3.4(1) \mu$s. (b) For nuclear spin 2.
$T_2^{\star} = 4.4(5)$ ms, simultaneously measured electron
$T_2^{\star} = 3.18(8) \mu$s. Fits are sine functions with a
decaying envelope $e^{(-t/T_2^\star)^\delta}$. Spin 1: $\delta =
2$, spin 2: $\delta = 1$. No readout correction.}
\label{Fig:T2Star} \vspace*{-0.2cm}
\end{figure}

We measure the nuclear dephasing time $T_2^{\star}$ by preparing
the nuclear spin in a superposition and the electron spin in
$m_s=0$, and let the system evolve for variable time. The electron
spin is then reset to $m_s=0$ before the nuclear spin is measured
along an axis that creates an effective detuning of approximately
1 kHz. We obtain $T_2^{\star} = 2.7(2)$ ms for nuclear spin 1 and
$T_2^{\star} = 4.4(5)$ ms for nuclear spin 2 (Fig.
\ref{Fig:T2Star}). An electron free-precession (Ramsey-type)
measurement is performed during the experiments (interleaved on a
$\mu{s}$ timescale), so that the electron and nuclear
$T_2^{\star}$ can be compared under the same magnetic field
fluctuations.

We expect the nuclear dephasing time to be set by a combination of
electron relaxation and magnetic field fluctuations (including the
nuclear spin bath). Electron relaxation gives a rate of
$\Gamma_{0,-1} + \Gamma_{0,+1} = 122(4)$ s$^{-1}$ (time constant
of 8.2 ms). To estimate the intrinsic nuclear dephasing timescale
$T_{2int}^\star$ we subtract the electron depolarization rate from
the inverse of the measured dephasing time $T_2^\star$. We find
$T_{2int}^\star \sim 4.0$ ms for spin 1 and $T_{2int}^\star \sim
9.5$ ms for spin 2. The difference inthese values could originate
from the differences in the nuclear spin's microscopic
environments.

\subsection{Magnetic field stability}

\begin{figure}[t]
\begin{center}
\includegraphics[scale=0.7]{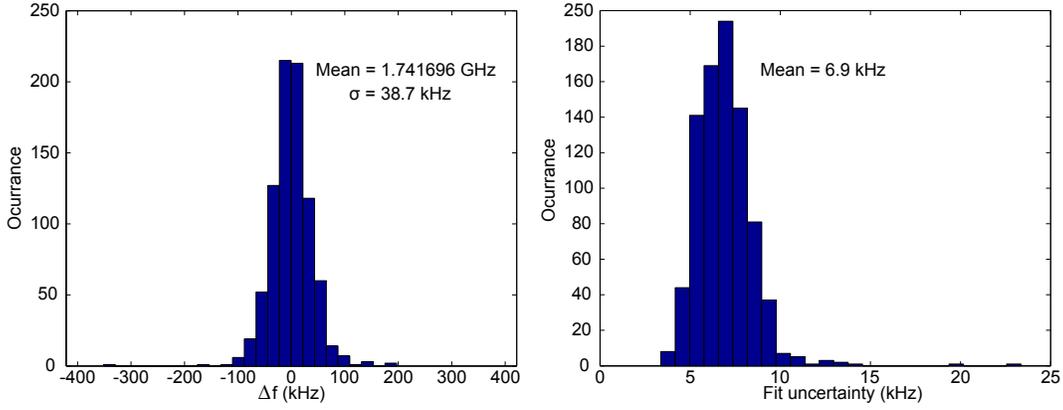}
\end{center}
\caption{\textbf{Magnetic field stabilization.} The magnetic field
during the error correction measurements was stabilized by a
feedback loop based on 840 measurements. (left) Electron energy
splitting. Fluctuations during the experiment of $38.7$ kHz
($1\sigma$) are observed, corresponding to $14$ mG. (right) The
average measurement uncertainty in a single instance of the
magnetic field measurement is $6.9$ kHz ($2$ mG).}
\label{Fig:B-Field_Stabilization} \vspace*{-0.2cm}
\end{figure}

We stabilize the magnetic field through a feedback loop by
periodically measuring the energy splitting of the NV centre. This
stabilization is required to counteract slow magnetic field drifts
(order of 0.1 G) over the measurement time. Figure
\ref{Fig:B-Field_Stabilization} characterizes the magnetic field
stability during the quantum-error-correction measurements with
simultaneous errors (taken over a total of $344$ hours, spread out
over 1 month). These values are representative for the other
measurements.

The measured residual slow fluctuations of the magnetic field
($38.7$ kHz, $0.014$ G) are small compared to the fast
fluctuations due to the $^{13}$C bath ($\sim 66$ kHz, $0.024$ G).
These slow fluctuations are expected to decrease the electron
$T_{2e}^\star$ from the instantaneous value $3.3\ \mu$s to $2.9\
\mu$s (and the nuclear $T_{2}^\star$ with approximately the same
factor).

Part of the fluctuations are caused by the uncertainty in the
measurements of the electron splitting. Figure
\ref{Fig:B-Field_Stabilization} shows that this effect is small,
because the measurement uncertainty of ($6.9$ kHz) is small
compared to the total drift observed ($38.7$ kHz).

\subsection{Nuclear $T_2$}

\begin{figure}[t]
\begin{center}
\includegraphics[scale=0.65]{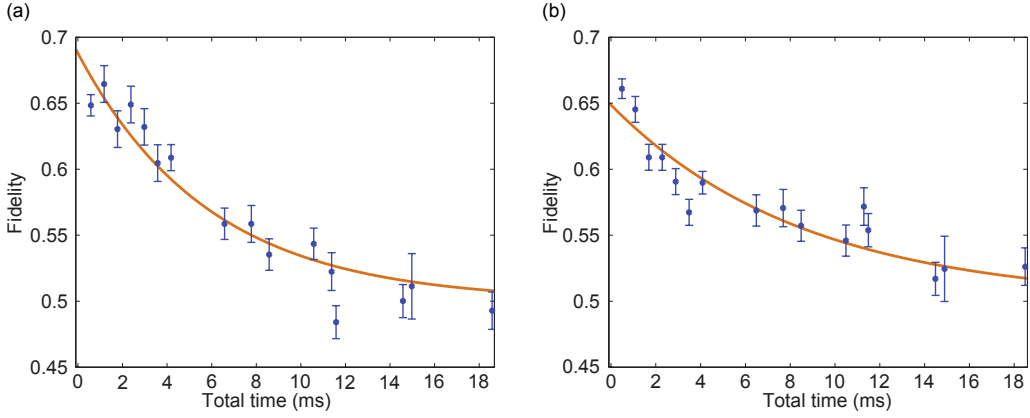}
\end{center}
\caption{\textbf{Nuclear spin echo experiments.} (a) for nuclear
spin 1: $T_2 = 5.9(8)$ ms. (b) for nuclear spin 2: $T_2 = 9(1)$
ms. Single exponential fits. No readout or initialization
correction.} \label{Fig:spin_echo} \vspace*{-0.2cm}
\end{figure}

The nuclear spin coherence times can be extended by decoupling
from the spin bath. Figure \ref{Fig:spin_echo} shows the results
of nuclear spin echo experiments. The required pi-pulse is
constructed in the same way as all nuclear gates in this work. The
electron is prepared in $m_s=0$ and re-initialized before the
pi-pulse, which makes it possible to use a conditional gate, and
re-initialized again to be used in the final measurement. We find
$T_2 = 5.9(8)$ ms (nuclear spin 1) and $T_2 = 9(1)$ ms (nuclear
spin 2).

\subsection{Nuclear dephasing $T_2^{\star}$ and depolarization $T_1$ under laser illumination}

Being able to re-initialize the electron spin without depolarizing
or dephasing the nuclear spins is essential for initializing the
multiqubit register and for performing partial measurements within
such registers.

\begin{figure}[t]
\begin{center}
\includegraphics[scale=0.7]{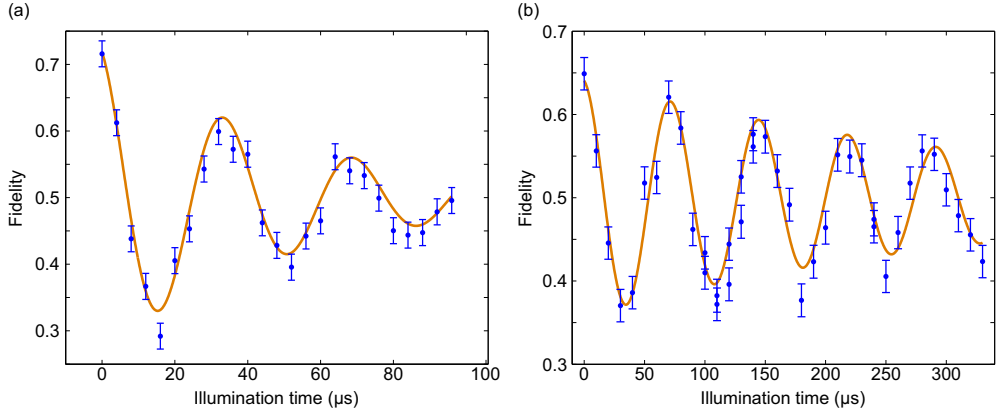}
\end{center}
\caption{\textbf{Nuclear Ramseys and $T_2^\star$ under
illumination.} (a) nuclear spin 1, $T_2^\star=51(7)$ $\mu$s. (b)
nuclear spin 2, $T_2^\star=0.35(9)$ ms. The laser power is the
same as used in the initialization, re-initialization and readout
steps. No readout or initialization correction.}
\label{Fig:T2Star_Laser} \vspace*{-0.2cm}
\end{figure}

Figure \ref{Fig:T2Star_Laser} shows a $T_2^\star$ under
illumination of $51(7)$ $\mu$s for nuclear spin 1 and $0.35(9)$ ms
for nuclear spin 2. These times are long compared to the time
required to re-initialize the electron spin ($\sim 2 \mu$s). For
example, for nuclear spin 1 this predicts a contrast loss of
approximately $1-e^{-2/50} = 0.04$.

\begin{figure}[t]
\begin{center}
\includegraphics[scale=0.7]{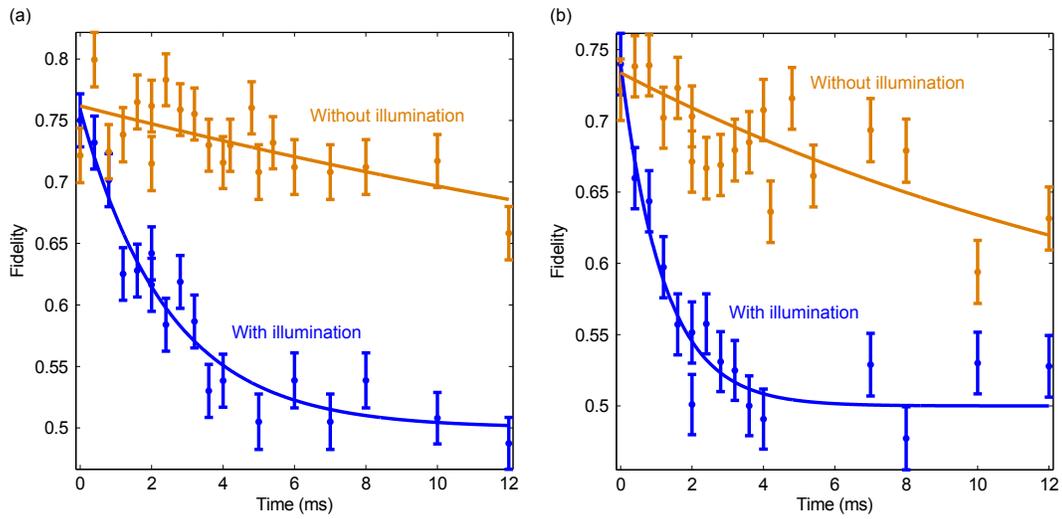}
\end{center}
\caption{\textbf{Nuclear T1 with and without illumination.} Lines
are exponential fits. (a) nuclear spin 1. (b) nuclear spin 2. The
laser power is the same as used in the initialization,
re-initialization and readout steps. No readout correction.}
\label{Fig:T1_Nuclei} \vspace*{-0.2cm}
\end{figure}

Figure \ref{Fig:T1_Nuclei} shows nuclear relaxation measurements
for the both spins, with and without laser illumination. The
nuclear spin is prepared in $\ket{0}$ and the electron spin in
$m_s=0$. We let the system relax for a variable time during which
the laser is either on or off. For the experiment without laser
illumination, the electron is reset by a short laser pulse ($2
\mu$s) so that it can be used to measure the nuclear spin state.
Without illumination, we find $T_1 = 0.04(1)$ s and $T_1 = 21(5)$
ms for spin 1 and 2 respectively. With illumination, we find $T_1
= 2.5(3)$ ms and $T_1 = 1.2(2)$ ms for spin 1 and 2 respectively.

The nuclear depolarization during laser illumination is slow
compared to the time it takes to re-initialize the electron spin
($\sim 2 \mu$s), so that the electron can be re-initialized
without depolarizing the nuclei. Note that the final signal
approaches a fidelity of $0.5$; prolonged laser light does not
create a preferential polarization for these nuclear spins.

\section{Fidelity estimates}

The estimates for the final fidelities for the three-qubit quantum
error correction protocol in the main text are obtained from the
values above as follows. We take the electron decoherence time
($T_{coh} = 2.86(4)$ ms) and the two nuclear spin intrinsic
dephasing times ($T_{2int}^\star \sim 4.0$ ms and $T_{2int}^\star
\sim 9.5$). As a rough estimate we approximate the three processes
for the three qubits by rates and add them to obtain a final decay
time $T_{est} = 1.4$ ms. The typical fidelity for the $1.8$ ms
quantum error correction protocol becomes $F_{est} =
e^{-1.8/T_{est}}/2 + 1/2 = 0.64$. This corresponds to estimated
process fidelity $F_{p,est} = 6F_{est}/4 -1/2 = 0.46$, similar to
the observed value. The average gate fidelity for the 10 nuclear
gates in the error correction protocol is estimated from
$F_{average} = 1/2\sqrt[10]{2F_{est}-1} + 1/2 = 0.94$.

\end{document}